\documentclass[longauth]{aa}
\usepackage[varg]{txfonts}

\usepackage[flushleft]{threeparttable}

\usepackage{graphicx}
\usepackage[unicode=true,
   pdfusetitle,
   bookmarks=true, 
   bookmarksnumbered=false, 
   bookmarksopen=false,
   breaklinks=false, 
   pdfborder={0 0 1}, 
   backref=false,
   colorlinks=false]{hyperref}

\usepackage{txfonts}
\usepackage[switch]{lineno} 
\usepackage{booktabs}
 
\usepackage[inline]{enumitem}

\usepackage{float}
\usepackage{natbib}
\bibpunct{(}{)}{;}{a}{}{,} % to follow the A&A style

\begin{document}

\title{The throughput calibration of the VERITAS telescopes}

   %\author{\href{https://orcid.org/0000-0002-8321-9168}{M. Nievas-Rosillo}
   %       \inst{1,2,3}%\thanks{something}
   %          \and
   %          \href{https://orcid.org/0000-0001-9868-4700}{G. Maier}\inst{3}\fnmsep%\thanks{something else}
   %       \\
    %          {\sc (The VERITAS Collaboration)}
    %      }

\author{
C.~B.~Adams\inst{1}\and
W.~Benbow\inst{2}\and
A.~Brill\inst{1}\and
J.~H.~Buckley\inst{3}\and
J.~L.~Christiansen\inst{4}\and
A.~Falcone\inst{5}\and
Q.~Feng\inst{6}\and
J.~P.~Finley\inst{7}\and
G.~M~Foote\inst{8}\and
L.~Fortson\inst{9}\and
A.~Furniss\inst{10}\and
C.~Giuri\inst{11}\and
D.~Hanna\inst{12}\and
T.~Hassan\inst{11,24}\and
O.~Hervet\inst{13}\and
J.~Holder\inst{8}\and
B.~Hona\inst{14}\and
T.~B.~Humensky\inst{1}\and
W.~Jin\inst{15}\and
P.~Kaaret\inst{16}\and
T.~K~Kleiner\inst{11}\and
S.~Kumar\inst{12}\and
M.~J.~Lang\inst{17}\and
M.~Lundy\inst{12}\and
G.~Maier\inst{11\and\footnotemark[2]}\and
P.~Moriarty\inst{17}\and
R.~Mukherjee\inst{6}\and
M.~Nievas~Rosillo\inst{11\and25\and26\and\footnotemark[1]}\and
S.~O'Brien\inst{12}\and
N.~Park\inst{18}\and
S.~Patel\inst{16}\and
K.~Pfrang\inst{11}\and
M.~Pohl\inst{19\and11}\and
R.~R.~Prado\inst{11}\and
E.~Pueschel\inst{11}\and
J.~Quinn\inst{20}\and
K.~Ragan\inst{12}\and
P.~T.~Reynolds\inst{21}\and
D.~Ribeiro\inst{1}\and
E.~Roache\inst{2}\and
J.~L.~Ryan\inst{22}\and
M.~Santander\inst{15}\and
A.~Weinstein\inst{23}\and
D.~A.~Williams\inst{13}\and
T.~J~Williamson\inst{8}
}

\institute{
Physics Department, Columbia University, New York, NY 10027, USA
\and
Center for Astrophysics $|$ Harvard \& Smithsonian, Cambridge, MA 02138, USA
\and
Department of Physics, Washington University, St. Louis, MO 63130, USA
\and
Physics Department, California Polytechnic State University, San Luis Obispo, CA 94307, USA
\and
Department of Astronomy and Astrophysics, 525 Davey Lab, Pennsylvania State University, University Park, PA 16802, USA
\and
Department of Physics and Astronomy, Barnard College, Columbia University, NY 10027, USA
\and
Department of Physics and Astronomy, Purdue University, West Lafayette, IN 47907, USA
\and
Department of Physics and Astronomy and the Bartol Research Institute, University of Delaware, Newark, DE 19716, USA
\and
School of Physics and Astronomy, University of Minnesota, Minneapolis, MN 55455, USA
\and
Department of Physics, California State University - East Bay, Hayward, CA 94542, USA
\and
DESY, Platanenallee 6, 15738 Zeuthen, Germany
\and
Physics Department, McGill University, Montreal, QC H3A 2T8, Canada
\and
Santa Cruz Institute for Particle Physics and Department of Physics, University of California, Santa Cruz, CA 95064, USA
\and
Department of Physics and Astronomy, University of Utah, Salt Lake City, UT 84112, USA
\and
Department of Physics and Astronomy, University of Alabama, Tuscaloosa, AL 35487, USA
\and
Department of Physics and Astronomy, University of Iowa, Van Allen Hall, Iowa City, IA 52242, USA
\and
School of Physics, National University of Ireland Galway, University Road, Galway, Ireland
\and
Department of Physics, Engineering Physics, and Astronomy, Queen's University, Kingston, ON K7L 3N6, Canada
\and
Institute of Physics and Astronomy, University of Potsdam, 14476 Potsdam-Golm, Germany
\and
School of Physics, University College Dublin, Belfield, Dublin 4, Ireland
\and
Department of Physical Sciences, Munster Technological University, Bishopstown, Cork, T12 P928, Ireland
\and
Department of Physics and Astronomy, University of California, Los Angeles, CA 90095, USA
\and
Department of Physics and Astronomy, Iowa State University, Ames, IA 50011, USA
\and
Centro de Investigaciones Energ\'eticas, Medioambientales y Tecnol\'ogicas (CIEMAT), Av. Complutense, 40, E-28040 Madrid, Spain
\and
Instituto de Astrof\'isica de Canarias, E-38205 La Laguna, Tenerife, Spain
\and
Universidad de La Laguna, Dept. Astrof\'isica, E-38206 La Laguna, Tenerife, Spain
%\and
%\nonumber$^*$) \email{mnievas@iac.es}
%\and
%\nonumber$^\dagger$) \email{gernot.maier@desy.de}
}
%   \institute{
%         Instituto de Astrof\'isica de Canarias, E-38205 La Laguna, Tenerife, Spain
%         \and
%         Universidad de La Laguna, Dept. Astrof\'isica, E-38206 La Laguna, Tenerife, Spain
%         \and
%         DESY, Platanenallee 6, 15738 Zeuthen, Germany\\
%         \email{mnievas@iac.es}\\
%         \email{gernot.maier@desy.de}
%    }

   %\date{Draft to submit to A\&A (2021, Sep 21)}

\date{Received <date> / Accepted <date>}

%\linenumbers
% \abstract{}{}{}{}{} 
% 5 {} token are mandatory
 
\abstract
{The response of imaging atmospheric Cherenkov telescopes to incident $\gamma$-ray-initiated showers in the atmosphere changes as the telescopes age due to exposure to light and weather. These aging processes affect the reconstructed energies of the events and $\gamma$-ray fluxes.} 
{This work discusses the implementation of signal calibration methods for the Very Energetic Radiation Imaging Telescope Array System (VERITAS) to account for changes in the optical throughput and detector performance over time.}
{The total throughput of a Cherenkov telescope is the product of camera-dependent factors, such as the photomultiplier tube gains and their quantum efficiencies, and the mirror reflectivity and Winston cone response to incoming radiation. This document summarizes different methods to determine how the camera gains and mirror reflectivity have evolved over time and how we can calibrate this changing throughput in reconstruction pipelines for imaging atmospheric Cherenkov telescopes.
The implementation is validated against seven years of observations with the VERITAS telescopes of the Crab Nebula, which is a reference object in very-high-energy astronomy.}
{Regular optical throughput monitoring and the corresponding signal calibrations are found to be critical for the reconstruction of extensive air shower images.
The proposed implementation is applied as a correction to the signals of the photomultiplier tubes in the telescope simulation to produce fine-tuned instrument response functions. This method is shown to be effective for calibrating the acquired $\gamma$-ray data and for recovering the correct energy of the events and photon fluxes. At the same time, it keeps the computational effort of generating Monte Carlo simulations for instrument response functions affordably low.}
{}


\maketitle
%________________________________________________________________

% Corresponding authors 
\renewcommand{\thefootnote}{\fnsymbol{footnote}}
\footnotetext[1]{\email{mnievas@iac.es}}
\footnotetext[2]{\email{gernot.maier@desy.de}}
\renewcommand{\thefootnote}{\arabic{footnote}}

%
%________________________________________________________________

\section{Introduction}\label{S:1}

When energetic $\gamma$-rays or charged particles (typically protons, atomic nuclei, or electrons) enter the atmosphere, they generate a cascade of secondary particles (also known as an extensive air shower or EAS) through pair production, bremsstrahlung emission, and, for the case of hadronic showers, fragmentation and decay of unstable mesons ($\pi^0$, $\pi^\pm$, and $K^\pm$). The resulting ultra-relativistic particles, traveling faster than the local speed of light in the atmosphere, produce coherent polarization of the dielectric medium. This in turn produces beamed Cherenkov radiation in the forward direction, forming a light pool at ground level of approximately $150\,\mathrm{m}$ radius. 

Imaging atmospheric Cherenkov telescopes (IACTs) sample this Cherenkov light pool where the optical assembly of each telescope collects and focuses the light onto a corresponding camera, comprising an array of photo-sensitive detectors. As the shower progresses from the top of the atmosphere to the ground, Cherenkov light is produced with a duration of up to 100 nanoseconds. 
From the ground, camera pixels register the flash of light to form an image of the shower. The shape and time-gradient of that image depends on the properties of the primary particle, on its arrival direction, and on the distance of the air shower to the telescopes. Images due to pure electromagnetic showers tend to be compact, well-defined, and of an approximately elliptical shape. Hadronic showers, on the other hand, tend to generate more penetrating secondary particles (e.g., muons) at higher transverse momenta, resulting in images with more clumpy shapes.
Particle shower images, particularly of $\gamma$-ray origin, can be described in terms of a first and second moment analysis whose parameters are used to derive the properties of the primary particle \citep{1985ICRC....3..445H}.

The observed spectrum of Cherenkov light that is generated in extensive air showers extends from about $200\,\mathrm{nm}$ to more than $700\,\mathrm{nm}$. The generated Cherenkov light  follows a $1/\lambda^2$ distribution; absorption and scattering of the light in the atmosphere affects mostly the ultraviolet (UV) part of the spectrum. As a result, the observed flux at ground level from $\gamma$-ray showers is significantly reduced below a wavelength of about $280\,\mathrm{nm}$. 
Even so, the bulk of the observed emission happens at short wavelengths, decreasing in intensity as the wavelength increases. This, in addition to the presence of strong airglow lines and increasing night sky background (NSB) starting from $550\,\mathrm{nm}$, explains why IACTs are designed to be mostly sensitive to blue and near-UV radiation.
The number of Cherenkov photons emitted in the shower is approximately proportional to the energy of the primary particle \citep{1985ICRC....3..445H,2015CRPhy..16..610D}. This is particularly true for air showers that are generated by primary electrons and $\gamma$-rays.

IACTs operate in harsh environments with variable weather conditions, wide temperature ranges, and occasional snow, rain, or dust storms. Nevertheless, these telescopes lack the protective buildings of other optical instruments, such as astronomical domes. The optical components of IACTs, which are designed to collect and focus the Cherenkov light, suffer from degradation processes due to the aforementioned weather conditions. In addition, the cameras of IACTs are usually made of photomultiplier tube (PMT) pixels operating at high voltages and they degrade as the total accumulated charge increases. During standard operations, PMTs are exposed to NSB light, typically inducing pixel currents of $5-10\,\mathrm{\mu A}$, and up to $15-20\,\mathrm{\mu A}$ during exceptionally bright moonlight conditions. These factors combined can induce aging that impacts both the optical and electronic response to incoming light, including Cherenkov light from EAS. 
Monitoring and correcting for the varying instrument performance therefore becomes a key element in the calibration and analysis of Cherenkov data.

\cite{Hillas_1990} originally proposed to monitor and calibrate IACTs using images generated from local high energy muons generated by hadronic showers. 
First implemented for the calibration of the Whipple telescopes \citep{1991ICRC....2..595F,1996APh.....5...27R}, various IACTs have used muon images to study the changing responses of the telescopes \citep{1994APh.....2....1V,2004APh....22..109A,2005AIPC..745..774M,2006PhDT.......205G, 2014arXiv1403.4550C}.
The Cherenkov emission generated by these particles is emitted in a cone with a roughly constant opening angle, appearing as a ring when observed from the ground at small incident angles. The radius of the ring depends on the properties of the incident particle (e.g., energy) and is not affected by instrument throughput. It can be compared with the number of photo-electrons seen by the telescopes to provide a measure of the efficiency of the telescope detecting optical light.
In practice, the analysis of muon images is challenging for many reasons. It involves a detailed geometrical reconstruction, particularly when part of the muon ring image falls outside the camera's field of view (FoV), when the shower trajectory is tilted or displaced with respect to the center of the FoV or if groups of pixels are malfunctioning. In addition, as Cherenkov photons from muons are produced close to the telescopes, atmospheric absorption is less severe and the Cherenkov photon spectrum is shifted to shorter wavelengths compared with photons generated by $\gamma$-ray showers \citep{2014arXiv1403.4550C}. Therefore, the calibration method using the muon images requires additional corrections or specific Monte Carlo (MC) simulations in order to provide an accurate estimation of the throughput for $\gamma$-ray showers \citep{2019ApJS..243...11G}. Due to this added complexity, the Very Energetic Radiation Imaging Telescope Array System (VERITAS) currently uses muons mostly as a supporting technique to monitor the calibration workflow (see section \ref{sec:calibration}).

This work discusses a different method to measure the total throughput. It is based on the characterization of mirror reflectivity and the continuous monitoring of camera gains. 
The relative variations of these parameters are used to correct the simulated $\gamma$-ray event pulse charges to provide throughput-corrected response functions. We discuss how this method was successfully implemented in VERITAS, providing a reliable way to determine the total optical and camera throughput, and to reconstruct shower energies and source fluxes.  

This document is structured as follows. Section \ref{sec:VTS} describes the VERITAS telescopes, giving a comprehensive overview of the main components that are used for both the standard operation of VERITAS and the calibration of the instrument. It also details the analysis techniques and workflows that are used to produce the standard science products of the telescope. The measurement of the total optical throughput and its evolution over time is presented in section \ref{sec:throughput_measurement}. In section \ref{sec:throughput_corrections} we show the implementation of throughput corrections, which are applied to the instrument MC simulations to account for instrument aging. The impact of the varying throughput on the performance of the instrument is evaluated in section \ref{sec:throughput_effects}, using metrics such as the energy threshold of the analysis, the differential sensitivity and the uncertainties in the reconstruction of fluxes. The validation of the method using real data collected over seven years is shown in section \ref{sec:validation_realdata}. Finally, we present a brief discussion of systematic uncertainties, limitations and possible improvements of the method in section \ref{sec:systematic_uncertainties}.

\section{The VERITAS telescopes}\label{sec:VTS}

VERITAS is a ground-based very high energy (VHE) instrument operating at the basecamp of the Fred Lawrence Whipple Observatory in southern Arizona, USA ($31^\circ 40' N$, $110^\circ57' W$, $1268\,\mathrm{m}$ elevation). 
It consists of four $12\,\mathrm{m}$ IACTs which use a shower imaging and moment analysis technique \citep{1996SSRv...75....1W} to detect $\gamma$-ray photons with energies above $85\,\mathrm{GeV}$. 

\begin{figure}[!htbp]
\centering\includegraphics[width=1\linewidth]{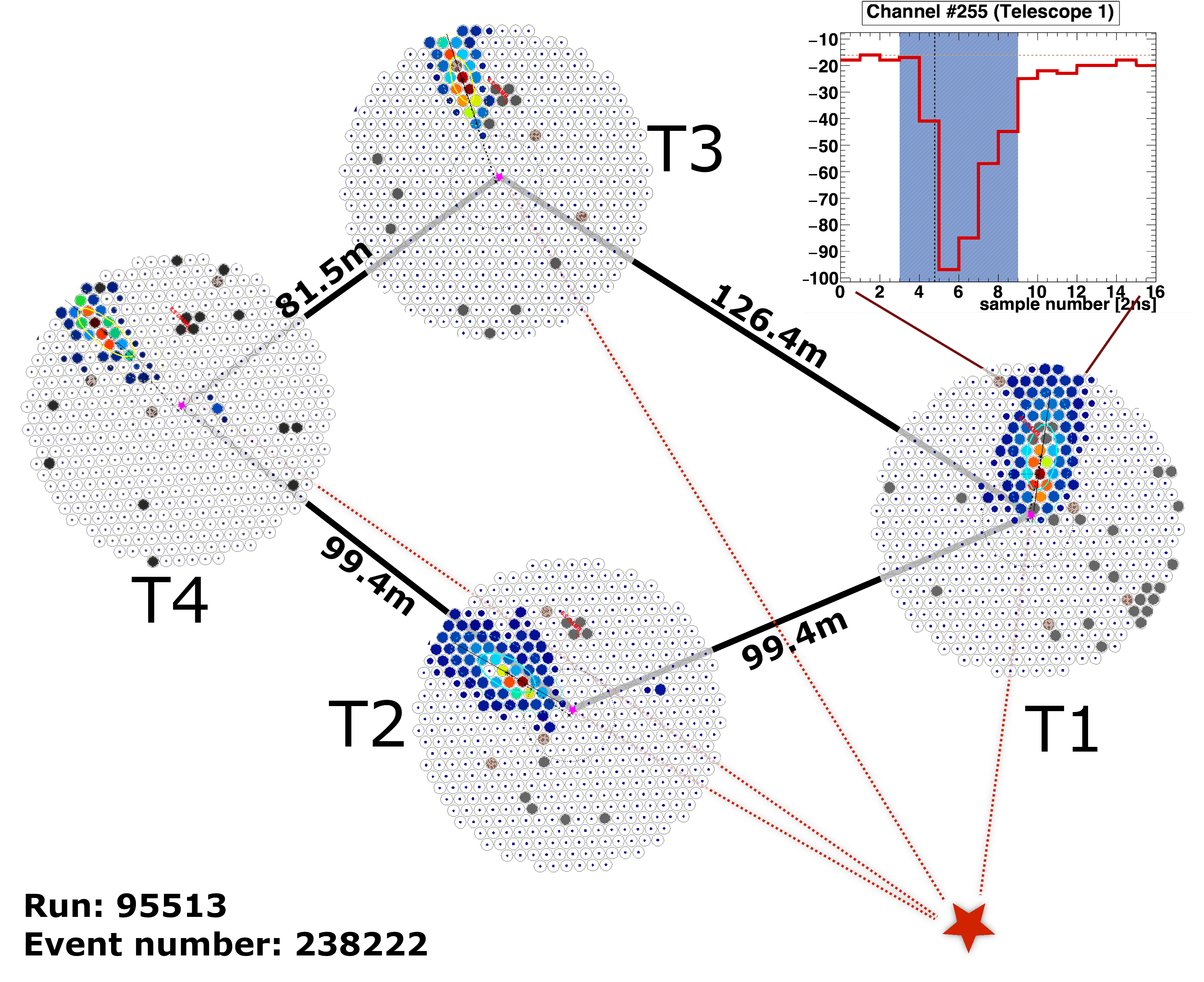}
\caption{Extensive air shower reconstructed by the VERITAS telescopes. The shower images for each camera have been integrated over time and cleaned using a two-level filter. Dead and disabled channels are shown in dark gray and brown respectively. The signal registered by each PMT is color-coded, with red colors representing higher signal than blue tones. 
An approximate geometrical reconstruction of the shower core location is illustrated with red dashed lines and a red star. The upper right plot shows the trace of one of PMT ({\tt \#255}) of T1 for reference. There, the signal registered for each sample of $2\,\mathrm{ns}$ is plotted in red and the integration window (six samples) is shown in shadowed blue.}
\label{fig:stereoscopic_shower}
\end{figure}

VERITAS operates in stereoscopic mode, with the array only triggering when a Cherenkov shower is detected by at least two telescopes. This allows a more accurate reconstruction of the shower properties (in particular, the incident direction and energy of the primary particle) while reducing the number of accidental triggers of a single telescope by NSB fluctuations or local muons. It also makes the discrimination of hadronic showers more efficient, which in turn boosts the sensitivity. An example of a stereoscopic reconstruction of a real EAS by VERITAS is shown in Figure \ref{fig:stereoscopic_shower}. 

The first VERITAS telescope (T1) started operations in February 2005 \citep{2006APh....25..391H}. Three more telescopes (T2, T3 and T4) were added in the following years. The full four-telescope array was commissioned by 2007 \citep{2007JPhCS..60...34K}. The observatory was subsequently upgraded to an optimized diamond-shaped array layout by moving T1 to its current position in 2009 \citep{2009arXiv0912.3841P}, and the camera and electronics were replaced with improved hardware and higher quantum efficiency (QE) PMTs in 2011-2012 \citep{2013ICRC...33.1124K}. 
With its current, post-upgrade, array configuration, VERITAS has been able to characterize $\gamma$-rays with energies from $\sim 85\,\mathrm{GeV}$ to $>30\,\mathrm{TeV}$, and can detect a point-like source with $\sim 1\%$ of the Crab Nebula flux in $25\,\mathrm{h}$.

\subsection{Telescope design}\label{sec:telescope_design}

The VERITAS telescopes follow a Davies-Cotton design \citep{1957SoEn....1...16D}.
The primary mirror of each telescope, hereafter also referred to as the ``dish,'' consists of 345 identical hexagonal facets. The resulting optical focal ratio of the system is $f/1$, for a focal length of $12\,\mathrm{m}$. Each facet is designed as a one-piece glass element with $\approx 61\,\mathrm{cm}$ sides \citep[][]{2008ICRC....3.1397R}. They are commercially ground and slumped for a radius of curvature of $23.97\pm0.01\,\mathrm{m}$ on average, and a spot size of $<10\,\mathrm{mm}$. 
The coating is made of aluminum, deposited at a rate of $3-8\,\mathrm{nm}/s$ under vacuum conditions with a purity better than $99.999\,\%$. Its total thickness is $180\,\mathrm{nm}$, the top $80\,\mathrm{nm}$ of which are oxidized during an anodizing process to improve durability and ensure a peak reflectivity of $92\,\%$ at about $320\,\mathrm{nm}$ \citep{2008ICRC....3.1397R}. 

Cherenkov light collected by the dish is focused onto the camera. In order to detect the brief  Cherenkov flashes, the cameras of the current configuration are composed of 499 high-quantum-efficiency  \citep{2011arXiv1110.4702N, 2011ICRC....9...14K} Hamamatsu R10560 PMTs located at the focal plane. Each PMT has a FoV of $0.15^\circ$ and is equipped with a Winston cone, a nonimaging device designed to minimize light collection gaps between the PMT photocathodes and limit the acceptance angle \citep{WINSTON197489,osti_7129508}, therefore reducing the background light intensity. Each Winston cone is made of plastic with an evaporated aluminum coating and a protective overlayer. This provides a reflectivity greater than $85\%$ above $260\,\mathrm{nm}$ \citep{2007arXiv0709.4517V}. 
After a careful optical alignment of the system \citep{2010APh....32..325M}, the on-axis point spread function (PSF) of the telescopes is about $0.10-0.12^\circ$ in diameter, leading to the ability to concentrate $80\%$ of the light within one pixel in the camera.
The optical PSF of the telescopes is frequently monitored, with observed variations of less than $0.02^\circ$ over long periods of time and for most elevations. Up to $0.6^\circ$ off-axis the degradation of the optical PSF with respect to on-axis observations is small, $\lesssim 0.02^\circ$, but it quickly increases at larger angles reaching $\sim 0.2^\circ$ at $1.2^\circ$ offset.
The total FoV of the telescopes is $3.5^\circ$. 

\subsection{Readout and trigger system}\label{sec:readout}

PMT signals are digitized using $500\,\mathrm{Msample/s}$ flash analog-to-digital converters (FADCs) with 8-bit dynamic range, capable of storing the waveforms in $64\,\mathrm{\mu s}$ memory buffers. 
VERITAS employs a three-level trigger. A first-level trigger (L1) or constant fraction discriminator (CFD) requires the PMT pulse height to be above a given threshold (typically 5-6 photo-electrons). A pattern trigger or telescope trigger (L2), requires L1 trigger signals to occur in three adjacent PMTs within the timing coincidence window. This pattern trigger is based on $400\,\mathrm{MHz}$ Xilinx Virtex-5 FPGAs for pixel neighbor coincidence, allowing triggers to occur before samples are read out. The time-aligning accuracy of this system is $\sim 0.2\,\mathrm{ns}$ and allows for a pixel-to-pixel coincidence window of $\sim 5\,\mathrm{ns}$. Finally, an array trigger (L3) requires a L2 trigger signal from at least two telescopes to happen, once corrected for the propagation time of the shower front across the array and the varying distance from each telescope to the central control building, within a $50\,\mathrm{ns}$ coincidence window.
A more detailed description of the VERITAS trigger system can be found in \cite{2013ICRC...33.3076Z}.

Coincidence window widths and CFD thresholds are optimized to trigger on low-energy gamma-ray showers while avoiding random coincident triggers from NSB. The optimization process consists of scans of the trigger thresholds when the instrument is exposed to NSB. The aim is to keep the L3 rates at a few hundred Hz, and to balance a low trigger energy threshold with avoiding data losses from dead-time. The typical dead-time for VERITAS, in its current configuration, is roughly $15\%$ for a data acquisition rate of about $300\,\mathrm{Hz}$.

\subsection{Data analysis}
\label{sec:VTS_performance}

VERITAS maintains two data analysis packages: Eventdisplay \citep{2017ICRC...35..747M} and VEGAS \citep{2008ICRC....3.1385C}. They allow independent reconstruction of the data, limiting the impact of systematic uncertainties due to the analysis software implementation on the scientific results. Each package performs a calibration of the signal collected for each shower by the four telescopes, a second-order moment analysis, and parametrization of the shower images. The resulting parameters are used to classify the showers as $\gamma$-ray-like showers or hadronic events. Using the stereo shower reconstruction, the arrival direction and the energy of the primary particle are estimated. Finally, from a comparison with reconstructed MC shower simulations, the effective collection area is evaluated and the events in excess of the estimated background are converted into high-level analysis products such as fluxes, light curves, and energy spectra. 
The results of the throughput analysis shown in this work were obtained using the Eventdisplay package, however they were also validated using VEGAS.

\subsubsection{Calibration}\label{sec:calibration}

The calibration process comprises the determination of the electronics baseline (pedestal) and its variation plus the measurement of the response of each individual PMT to incident light, that is absolute and relative gain differences between PMTs.
FADC inputs are AC coupled and a DC voltage offset is added to the signal inputs (pedestal) so that positive and negative fluctuations around the mean value (pedestal variation) due to NSB variations can be measured. Artificially triggered pedestal events are injected during observation runs with a frequency of $1\,\mathrm{Hz}$ and used for the estimation of pedestal level and variation. The sky brightness and therefore the background light level might change during observations; pedestal levels and variations are consequently updated every three minutes.

PMT gains are monitored and calibrated by using a flasher light source. The VERITAS FADCs have two gain channels for a wide dynamic range of the readout chain \citep{2005ICRC....5..379H}. The calibration software reconstructs the relative gain of the pixels, timing differences (e.g., due to differences in the cable length for each channel), and relative calibration of the high voltage settings and the high- and low-gain readout channels by using uniformly illuminated camera events generated with flasher light sources \citep{2010NIMPA.612..278H}. 
Each flasher unit consist of blue ($370\,\mathrm{nm}$) light-emitting diodes (LEDs), driver electronics, and a front face made of an opal diffuser which spreads the light from the LEDs and distributes it almost homogeneously across the entire PMT camera. The flasher pulses span eight brightness levels (upgraded to fifteen in 2016), covering the dynamic range of VERITAS for both high- and low-gain readout. 
Absolute gain calibration is determined on a periodic basis following the procedure described in \ref{sec:gains}. 
Relative gain differences between PMTs are monitored daily, and corrected during the data analysis. The inter-calibration between the two gain channels is performed on a monthly basis by recording calibration runs with a particularly long readout window. This is needed to avoid truncation and provide samples at the end of the trace that allow us to estimate the low-gain pedestal. For each run, half the camera is operated at a reduced gain to force it to stay in high-gain mode, while the other half operates in low-gain.

Finally, the results of the analysis of Cherenkov light from muons measured by the VERITAS telescopes \citep{2013ICRC...33.3096T} are used to monitor the calibration, in conjunction with the procedures described below (single photo-electron runs, optical PSF, and mirror reflectivity measurements). This approach mitigates several limitations of a calibration based on muons only, especially the differences in the wavelength spectrum between Cherenkov light from muons and from air shower events.

\subsubsection{Signal extraction and image analysis}\label{sec:signal_extraction}

Signals are extracted from the FADC traces using a two-pass integration method. 
The first pass uses a long integration window to determine the pulse arrival time and to derive the optimal position of the time integration window, using a linear fit of the pulse arrival times along the image axis  \citep{2006APh....25..391H}. A second pass performs the trace integration, using a short window of typically six samples ($12\,\mathrm{ns}$).
An example of a PMT trace for a real event and the optimized integration window is shown in Figure \ref{fig:stereoscopic_shower}.

A two-level filter is then used to extract the signal pixels and clean the image. First, the brightest parts of the shower image (core pixels) are localized at five times the pedestal root mean square (RMS). If a Gaussian distribution of pixel signals is assumed for images containing only noise, this would correspond to a surviving rate of less than one in $3\times10^6$. This is enough to remove most clusters of pixels containing only noise that might exist in the full image. Then, to avoid cutting out the edges of the shower images, adjacent (boundary) pixels with at least 2.5 times the pedestal RMS are added to the core pixels.
The resulting shower image, which resembles an ellipse in the case of $\gamma$-rays, is parameterized using a second moment analysis, based on the method originally developed by \cite{1985ICRC....3..445H}. 

Images from showers with large inclinations, large offsets, or originating from high energy showers may not be fully contained in the camera. These ``leaking events'' cause a loss of signal, degrade energy resolution and take a toll on the performance. Using a log-likelihood fitting algorithm, some of these showers are recovered \citep{2017ICRC...35..747M}, particularly improving the performance of the array at high energies.

\subsubsection{Direction, shower core, and energy reconstruction}\label{sec:direction_energy}

The direction of the primary particle and the impact parameter of the shower core on the ground are derived using stereoscopic techniques \citep{1999APh....12..135H, 2017ICRC...35..747M}.
The energy of each $\gamma$-ray event is estimated from a comparison using lookup tables built with simulations.
The lookup tables encode the dependency of the distance of the air shower core to each telescope, integrated signal of the shower image (size), noise level, azimuth, zenith angle and array configuration.

\subsubsection{Gamma-hadron separation}\label{sec:gamma_hadron_sep}

Cosmic rays represent the largest fraction of the data for all $\gamma$-ray observations\footnote{A notable exception are flaring events such as GRBs, for which the rate of $\gamma$-rays may exceed the rate of background events for the first few seconds after analysis cuts have been applied \citep{2020arXiv200607249M}.}. In order to separate signal from background cosmic-ray events, it is common practice to compare the shape of observed shower images with simulated $\gamma$-ray showers. 
To break the dependency of length and width of the image ellipse on the energy of each event, Eventdisplay uses the median reduced scaled parameters (MSCP) of width (MSCW) and length (MSCL) \citep{2006APh....25..380K}:

\begin{equation}
\mathrm{MSCP} = \frac{1}{N_{tel}} \sum_{i=1}^{N_{tel}}{\frac{p_i - {\bar p}_{MC} (\theta, \phi, \alpha, \mathrm{NSB}, \mathrm{ATM}, \mathrm{size}, r)}{\sigma_{90}}} ,
\end{equation}

where $p_i$ is the measured parameter (in our case width or length) for telescope $i$, $\bar p_{MC}$ and $\sigma_{90}$ the expected value and width of the distribution for $90\%$ of the events, which are read from lookup tables that are filled using simulated gamma rays. Finally, $\theta$ is the zenith angle, $\phi$ the azimuth angle, $\alpha$ the offset angle of the telescope pointing direction with respect to the nominal source position, NSB the night sky background, ATM the atmospheric profile, size the integrated charge of the shower image, and $r$ the distance of the shower core to  telescope $i$. VEGAS, in contrast, uses nonreduced versions of these parameters, centered at 1 instead of 0.

The separation of $\gamma$-rays from background cosmic-ray events is done using boosted decision trees (BDTs) \citep{2017APh....89....1K}. BDTs are implemented as part of the Toolkit for Multivariate Data Analysis (TMVA) package \citep{2007physics...3039H} and trained on a set of simulated $\gamma$-ray and observed hadronic showers using reconstructed shower parameters: MSCW, MSCL, shower core position, height of the maximum of the Cherenkov emission, and quality of the energy reconstruction.

With different science goals in mind, three sets of cuts are usually defined in VERITAS, each requiring different BDT scores and different minimum image brightness. ``Soft cuts'' provide a low energy threshold but larger acceptance of background events. They are ideal for sources with very steep energy spectra which emit large fluxes from tens of GeV to a few hundreds of GeV. ``Moderate cuts'' are the most widely used and are adequate for most analyses, as they provide a good balance between energy threshold and sensitivity. ``Hard cuts'', particularly combined with higher multiplicity (e.g., 3-telescopes) provide the best background rejection. Though the energy threshold is high ($\gtrsim 300\,\mathrm{GeV}$), this set of cuts is used when the source of interest is weak but with emission extending well into the TeV energy range. Unless otherwise stated, ``moderate cuts'' have been used throughout this work.

\subsubsection{Spatial and spectral reconstruction}\label{sec:spatial_spectral}

Event counts are converted into energy spectra taking into account the effective areas of the telescopes and dead-time corrected exposure time \citep{2008ApJ...679.1427A}.
The background at each point in the sky is estimated using either the reflected region method or the ring-background model approach \citep{2007A&A...466.1219B}.

\subsection{Monte Carlo simulation}

Instrument response functions (IRFs), required for both the separation of $\gamma$-rays from background events and for the reconstruction of energy spectra and light curves, are generated using large sets of simulations of $\gamma$-ray showers.

The propagation of extensive air showers in the atmosphere is carried out using the CORSIKA package \citep{1998cmcc.book.....H}
taking into account all relevant particle interactions.
The generation, propagation, scattering, and absorption of Cherenkov light is simulated using measurements of molecular and aerosol content of the atmosphere from a radiosonde (station number {\tt \#72274}) which operates from Tucson, Arizona, approximately $60\,\mathrm{km}$ away from the VERITAS site. From the radiosonde data, two sets of simulations are generated, the first with an average winter atmospheric profile and the second for summer atmosphere.

The response of the instrument is simulated using the GrOptics\footnote{\protect\url{https://github.com/groptics/GrOptics}} and CARE\footnote{\protect\url{https://github.com/nepomukotte/CARE}} packages for the optical and camera simulations.
The optical simulations take into account the properties of the primary mirrors including facet alignment, reflectivity, diffuse scattering and all relevant shadowing elements of the telescopes.
The photo-detector and electronic simulation code CARE is used to model the  digitization, trigger, and readout processes including noise components.
The MC model is validated through extensive comparisons with calibration measurements.

The generation of MC simulations for instrument response functions is an important computational effort, given the large parameter space that needs to be covered.
The MC instrument model for the array configuration after the upgrade of the cameras was generated to accurately reproduce the array performance for the period October 2012 - June 2013.

\section{Throughput measurements}\label{sec:throughput_measurement}

The collection efficiency of Cherenkov light at ground level by the telescopes and its conversion to measurable signals, that is telescope throughput, strongly depends on the properties of the cameras and dishes, which change over time.
At the camera, PMT gains and the QE affect the conversion of photons into charge pulses in each pixel. Similarly, the combined effect of the reflectivity of the mirror facets and the collection efficiency of the Winston cones affects the total optical light collected at each PMT photocatode. 
Furthermore, some of these parameters vary on different time scales and some are wavelength dependent, turning the characterization of the different parts of the instrument into an important and very challenging step in the analysis of data produced by IACTs. 
This section describes how these components are studied in VERITAS.

\subsection{Camera gains and gain factors}\label{sec:gains}

The current detector model of VERITAS assumes an average value of the absolute PMT gain $G_{i,MC}$ for each telescope $i$, estimated after the upgrade of the PMT cameras in 2011-2012:

\begin{itemize}
\item $G_{1,MC} = 5.38\,\mathrm{dc/pe}$
\item $G_{2,MC} = 5.29\,\mathrm{dc/pe}$
\item $G_{3,MC} = 5.29\,\mathrm{dc/pe}$
\item $G_{4,MC} = 5.73\,\mathrm{dc/pe}$
\end{itemize}

\noindent where dc stands for digital counts and pe for photo-electrons.

In practice, the individual absolute gains are slightly different for every PMT and they vary over time due to temperature fluctuations. Supply voltage fluctuations are are kept stable at the subvolt scale, therefore are unlikely to have a large impact in the absolute gains.

At the beginning of each observing campaign, usually in October, absolute and relative gains are measured and adjusted by changing the HV settings of the telescopes. The process is usually repeated once or twice during the season. This brings the gains closer to the nominal values shown before and results in absolute gain swings of, at most, $10\,\%$ over months. The resulting PMT charge distribution is also optimized to have $\lesssim 5\%$ width. Part of the aim of this work is to reduce the impact on the IRFs of the remaining gain variations left after the flat fielding process.

Gains are measured in VERITAS in two independent ways, each with its own set of assumptions. 
The first approach directly measures the absolute gain by detecting the signal from single photo-electrons. This is achieved by placing a cover to reduce the optical transmission in front of the camera. 
The second method evaluates the absolute gains of the PMTs statistically, using ``LED flashers'' to uniformly illuminate the camera. Flasher runs are collected daily to perform a relative gain correction directly in the analysis chain.

\subsubsection{Single photo-electron (SPE) runs}

Weak, pulsating sources of light can be used to study the response of PMTs to a single photo-electron. The measured charge from a single photo-electron is, on average, proportional to the number of electrons produced by the PMT. The proportionality constant is the gain of the system, $G$ 
\citep[see e.g., ][]{2011ICRC....9...14K,2010NIMPA.612..278H,macleod2007measuring}.

In VERITAS, single photo-electron light levels are achieved by attenuating the light of the flasher with 
a custom-made camera cover with a small hole aligned with the center of each PMT \citep{2008ICRC....3.1417H}. Each of these holes allows only about $1.6\%$ of the light to pass. With such a device, not only are the required low-illumination conditions achieved, but NSB is suppressed by the same amount.

A histogram of the accumulated charge shows a series of peaks. The first peak describes the pedestal, the second the mean value of the SPE charge in digital counts. The separation depends on the absolute gain set by the HV settings used during the data acquisition, making it possible to derive the gain-voltage relationship and measure the absolute gain.
Since the absolute gain can locally be approximated as having a linear dependence on voltage, VERITAS takes SPE runs at both nominal and a slightly increased voltage ($110\%$) so that the two peaks are better separated.

The shape of the single-electron peak follows a Polya distribution \citep{2018APh...102...56L}, which is a special case of a negative-binomial distribution. The comparison to photo-statistic gains, described in the next section, requires a correction factor for each pixel, which quantifies the deviation from Poisson statistics. 
SPE runs are taken in VERITAS approximately once a month, as they require the manual installation of the custom cover on each of the telescope cameras, temporarily interrupting the data taking and leading to increased observer fatigue and safety risks.

\subsubsection{Photo-statistic gains from nightly flasher runs}

The second method to measure absolute gains in VERITAS uses nightly flasher runs. We refer to these gains as photo-statistic gains. During a flasher run, the mean charge of a pulse on a PMT, $\mu$, is statistically proportional to the mean number of photo-electrons at the first dynode, $N_{pe}$:

\begin{equation}
\mu = G \times N_{pe} .
\end{equation}

The proportionality constant is the absolute gain $G$. It can be determined by assuming that $N_{pe}$ follows Poisson statistics, which implies that the variance in photo-electrons is approximately the mean charge $\mu$, hence $\sigma_{pe} \simeq \sqrt{N_{pe}}$. After the dynodes and preamplifier have amplified the signal, the variance becomes 

\begin{equation}
\sigma^2 \simeq G^2 \times N_{pe} = G \times \mu .
\end{equation}

Fitting a linear function to the charge variance $\sigma^2$ over $\mu$ provides the gain as the slope and the variance of the pedestal noise as the intercept.

\begin{figure}[!htbp]
\centering\includegraphics[width=1\linewidth]{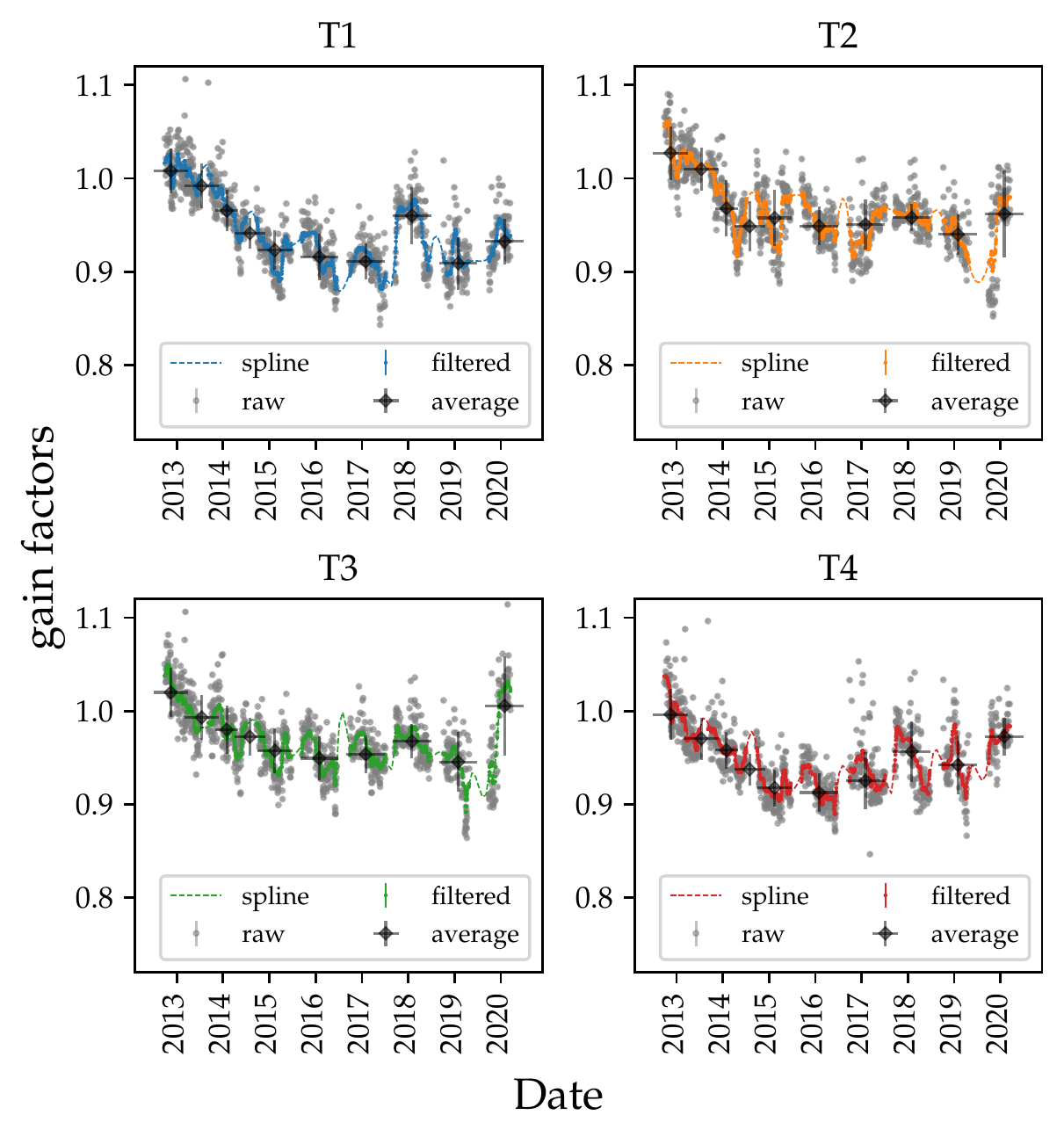}
\caption{Time-dependent changes of photo-statistic gains normalized to the nominal absolute gains used in the baseline MC model for the current detector model. This detector model was constructed in 2012, and reproduces the characteristics of VERITAS after the PMT camera upgrade. Black points show the average gain factors for each instrument epoch. Gray points show the individual unfiltered gain factor values per telescope and daily flasher run. Blue, orange, green and red points show the result of a median filter of the individual gain factors. Curves of the same colors represent a spline interpolation of the filtered values. Even though spline interpolation can locally diverge when there is no data available, we note that this occurs only during Summer breaks, where no data is taken in VERITAS, therefore it is not a concern for this study. Error bars represent statistical uncertainties.}
\label{fig:photostat_gfactors}
\end{figure}

\subsubsection{Comparison of absolute gain measurement methods}

The two methods of measuring the gain agree within $5-6\%$, which is assumed to be the systematic uncertainty for the absolute gain estimation. Photo-statistic gains can be determined using runs that are taken once a night, therefore allowing better monitoring. We consequently opted to use them for our throughput calibration purposes, as opposed to SPE gains. 

So far we have only discussed the evolution of the gains, but the QE of the PMTs may also change as the detector ages. 
The possible aging of VERITAS PMTs was covered in \cite{2016JInst..1111015G}. In that work, a set of 20 PMTs were removed from the telescopes and compared to 20 spare PMTs that were never operated and therefore are expected to exhibit their original properties. The absolute quantum efficiency was measured to be consistent for both samples at the wavelength range in which VERITAS detects Cherenkov radiation from $\gamma$-ray showers. More recently, during the Summer break of 2020, another aging study was done using the same sets of PMTs. This time, the test consisted of the determination of the HV-dependence of PMT gains, which is related to their QE. Again the properties of both sets of PMTs were found to be compatible within uncertainties.

\subsubsection{Gain factors}

Once the average absolute camera gains ($G_i$) are measured for each telescope $i$, they can be compared with the gains assumed in the reference MC simulations ($G_{i,MC}$) to obtain the gain factors ($g_i = G_i / G_{i,MC}$) needed to correct the simulated signals to account for changes in the camera gains. Figure \ref{fig:photostat_gfactors} shows the gain factors as derived from photostatic gains.

\subsection{Optical throughput and mirror reflectivity}

VERITAS is located in southern Arizona, where at least three months during summer are under direct influence of the North American monsoon, in addition to the all year round varying weather conditions. The mirror facets of the telescopes are consequently affected by mechanical and chemical degradation of their reflecting surfaces (e.g., due to fine dust grains which scratch the reflective coating, pollutants chemically reacting with the mirror materials, or temperature oscillations inducing deformations on the underlying mirror substrate). 
Furthermore, periodic mirror cleaning, re-coating, and Winston cone cleaning contribute to partially recover the reflectivity of VERITAS telescopes, but it also adds an additional variability component to the optical performance of the telescopes.

The degradation of the mirrors changes their reflective properties. The total reflectivity is the combination of diffuse and specular reflectivity. The first is the tendency of a surface to reflect the incident radiation in random directions. Degraded mirror surfaces can cause high diffuse reflectivity, scattering photons and making the mirrors unable to focus and properly form an image. For imaging instruments, diffuse reflectivity is undesirable as it increases background photon noise at the focal plane and degrades the optical PSF. Specular reflectivity refers to the ability to form an image of a given object at the focal plane. It is very sensitive to mirror aging since it depends on the material of the substrate, how accurately the mirror is figured, how smooth its surface is and the type of coating. 

Mirror coating degradation causes a wavelength-dependent effect on the optical throughput of the instrument \citep{app10217538}, with reflectivity at short-wavelengths more severely affected than that for longer wavelengths. This is relevant for IACTs because the bulk of the Cherenkov emission concentrates in the near-UV and  blue part of the visible spectrum. 

The optical properties of the Winston cones are not expected to vary dramatically over time as they are protected from the elements and the camera is fully closed while the telescopes are stowed (i.e., during daylight and during bad weather conditions). Nonetheless, these components were examined over time and no evidence of degradation in their light collection efficiency was detected. Therefore, we  concentrate on the reflectivity of the telescope dishes and treat Winston cone efficiency as a source of systematic uncertainty.

\subsubsection{Laboratory measurements of specular reflectivity}\label{sec:reflectivity_lab}

Beginning with the VERITAS inauguration in 2007 and continuing until 2015, fifteen facets from different parts of the dish (top, middle, bottom) were regularly removed 
to monitor the changes of the wavelength-dependent reflectivity over time. 
This was done in the laboratory using a broad-spectrum light source, an adjustable filter wheel, and a photometer (Oriel 71610). 
Measurements were compared to a calibration mirror made of pure aluminum, periodically re-coated to ensure consistency in the measurements \citep{roache2007}.

Even though it was intented to measure specular reflectivity, this method had the disadvantage of not being able to fully discriminate between diffuse and specular reflectivity as it relied on a photometer and not on a real imaging detector.
It also did not evaluate the impact of PSF changes on the performance of the instrument collecting Cherenkov light. 
Finally, the evolution of the reflectivity of a few mirrors was measured, rather than that for the entire dish. 

Laboratory measurements of specular reflectivity were dropped in 2016 as the achieved accuracy was limited by the complexity of determining  the ratio of specular to diffuse reflectivity.

\subsubsection{Whole dish reflectivity (WDR)}\label{sec:WDR}

The WDR method was initially developed by MAGIC \citep{2007APh....27..509M} and after a few preliminary tests in 2010 and 2011, the method was adopted by the VERITAS observatory starting from 2014 \citep{Archambault:2013zja, 2021APh...12802556A}.
The WDR technique uses a wide-field CCD camera attached to the dish, near its center, which is pointed directly toward a target at the focal plane in front of the PMT camera. 
In order to reproduce the response of the mirrors to Cherenkov light combined with the QE of the PMT, which peaks at $\sim 340\,\mathrm{nm}$, the CCD (SBIG ST402ME) is equipped with a blue filter from the same manufacturer (Diffraction Limited$^\copyright$). This effectively shifts the response of the CCD camera to shorter wavelengths and provides a better match to the spectral response of the PMT camera to extended air showers. 
The target is made of Spectralon, a fluoropolymer which exhibits Lambertian behavior (i.e., high diffuse reflectivity). This ensures that the apparent brightness is the same regardless of the observing angle. The reflectivity of the Spectralon plates used in VERITAS has been checked over the years and no significant change could be detected, setting a conservative limit of $\lesssim 5\%$ to their evolution, well within the total systematic uncertainty of the reflectivity determination. With this set-up, the CCD camera can be used to measure simultaneously the brightness of a reference star in the FoV and its reflection generated by the dish on the Spectralon. 
Flat and dark frames are used to correct for effects such as vignetting or pixel-to-pixel response, and to subtract the electronic dark signal in the image. The comparison of the brightness of the reference star and its reflected spot yields an estimation of the total reflectivity of the dish. 
The advantage of this method resides on its simplicity and robustness, using just a standard CCD with a color filter and target plate over the focal plane. Being based on differential photometry methods, it is insensitive to changes in the QE or the gain of the CCD itself. The main disadvantage of the WDR method is that the employed CCD is blind to UV radiation and therefore the spectral averaged reflectivity does not fully match the reflectivity for Cherenkov light.

\subsubsection{Optical throughput and optical throughput factors}

In a manner similar to the case of the gains, this study focuses on the relative changes of the reflectivity with respect to what is simulated for each telescope $i$ in the detector model. From the two possible methods to estimate the reflectivity of VERITAS mirrors, we adopted the WDR method described in section \ref{sec:WDR}.

VERITAS simulations take into account the complete spectral response of each telescope as a function of wavelength (see Appendix \ref{sec:appendix_reference_reflectivity}). On the other hand, the WDR method only provides an averaged reflectivity for each telescope $i$, measured in the blue part of the visible spectrum. The calibration of the optical throughput requires the estimation of the same wavelength-averaged reflectivity for the reflector model of the current instrument configuration $R_{i,MC}$ and the assumption that the differences due to the spectral mismatch that exists between the Cherenkov spectrum and the assumed blue filter are small.

The spectral average is determined assuming a standard Bessel-B filter \cite{1990PASP..102.1181B}, similar to the one used with our CCD. The transmission of this filter is multiplied by the reference reflectivity curves from the reflector model. The resulting curve is integrated to obtain the average reflectivity $R_{i,MC}$ for each telescope:

\begin{itemize}
    \item $R_{1,MC} = 0.828$
    \item $R_{2,MC} = 0.841$
    \item $R_{3,MC} = 0.852$
    \item $R_{4,MC} = 0.846$
\end{itemize}

In order to estimate the collection efficiency error due to the use of a B-filter and a CCD to characterize the response of the telescope to Cherenkov light, we computed the wavelength-averaged reflectivities, this time convolving each mirror reflectivity curve with a typical Cherenkov spectrum from a $\gamma$-ray induced shower. The resulting average reflectivities are between $0.79$ and $0.82$ for all telescopes. The differences are therefore well within the assumed systematic uncertainty on the mirror reflectivity ($\lesssim 10\,\%$).

\begin{figure}[!htbp]
\centering\includegraphics[width=1\linewidth]{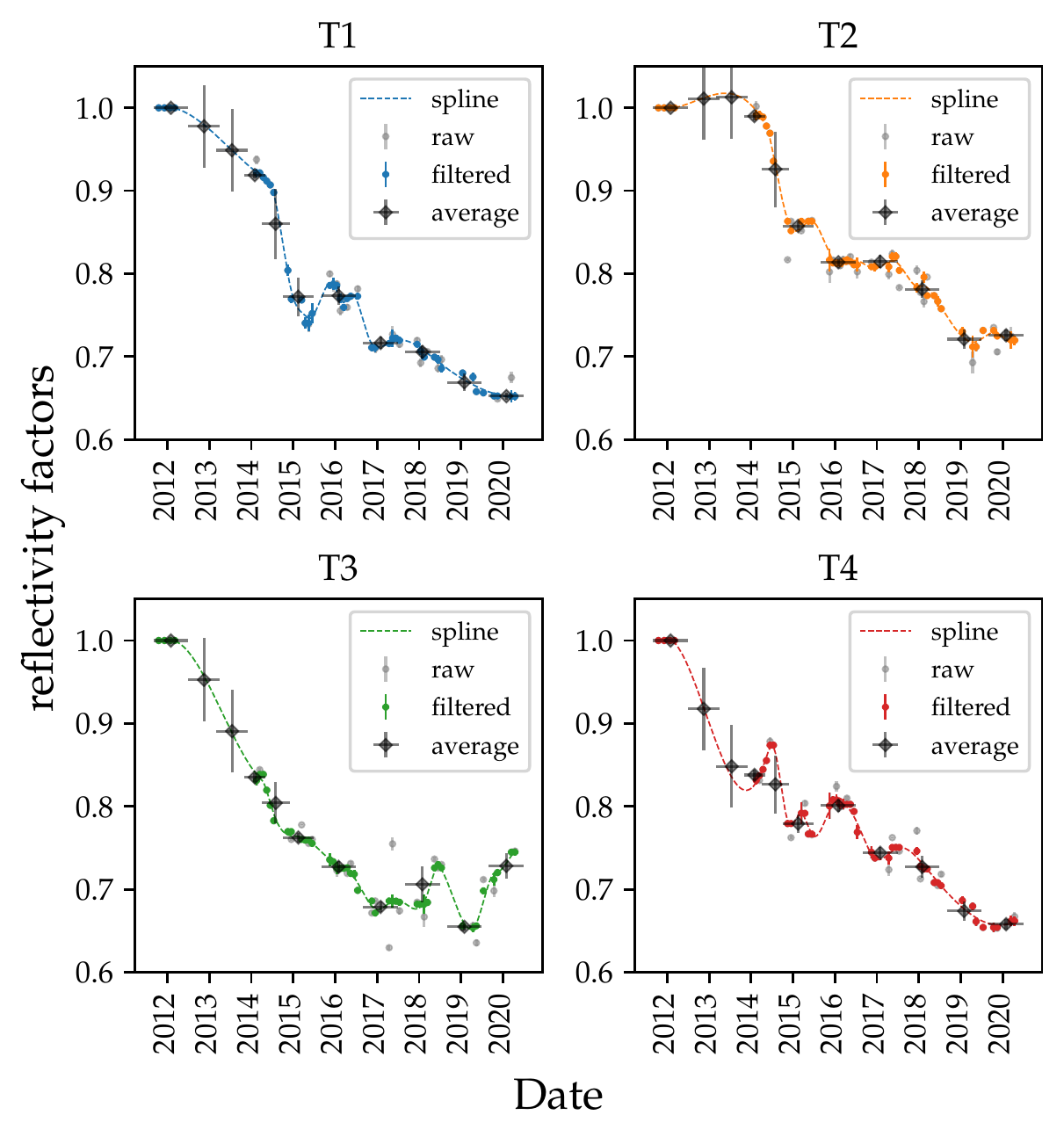}
\caption{Time-dependent changes of the reflectivity factors obtained from WDR measurements. Black points show the average reflectivity factors for each instrument epoch. Blue, orange, green, and red points show the individual reflectivity factor measurements for each telescope. Curves of the same colors represent a spline interpolation, which removes outliers. The first point (early 2012) is extracted from the simulations and it serves as the reference for the calculation of the reflectivity factors, therefore adopting a value of 1 as reflectivity factor. We note that a value of the reflectivity factor slightly greater than 1 just means that the reflectivity at that time was slightly higher than at the time where simulations were produced.}
\label{fig:wdr_tfactors}
\end{figure}

Calculated based on the reflectivity measurements, the optical throughput factors or reflectivity factors can be defined as $r_i = R_i / R_{i,MC}$. 
Their evolution over time is shown in Figure \ref{fig:wdr_tfactors}. 
These reflectivity factors are needed to correct our MC simulations so that they reflect the changes in the mirror reflectivity. 
The first point is fixed at unity as it represents the reflectivity values actually used in the simulations. Because the whole-dish reflectivity only started to be measured routinely in 2014, the second and third points of each panel of Figure \ref{fig:wdr_tfactors} had to be interpolated. 
We artificially assigned them a comparatively larger uncertainty of $\pm 5\%$ in absolute reflectivity ($\pm 0.05$ in the plot). 
As can be seen from the figure, reflectivity quickly degraded by $20-30\%$ from 2013 to 2015, coinciding with a period of reduced cadence of re-coating degraded mirror facets. 
Since 2015, the telescope reflectivity continues to decrease, but at a much slower pace. Improvements in the reflectivity are also present in our data for specific times, coinciding with cleaning and re-coating of the most damaged mirror facets. This process usually involves the exchange of $\sim 100$ mirror facets, and its effects are smoothed in the spline fitting case.

\subsection{Total throughput}

\begin{figure}[!htbp]
\centering\includegraphics[width=1\linewidth]{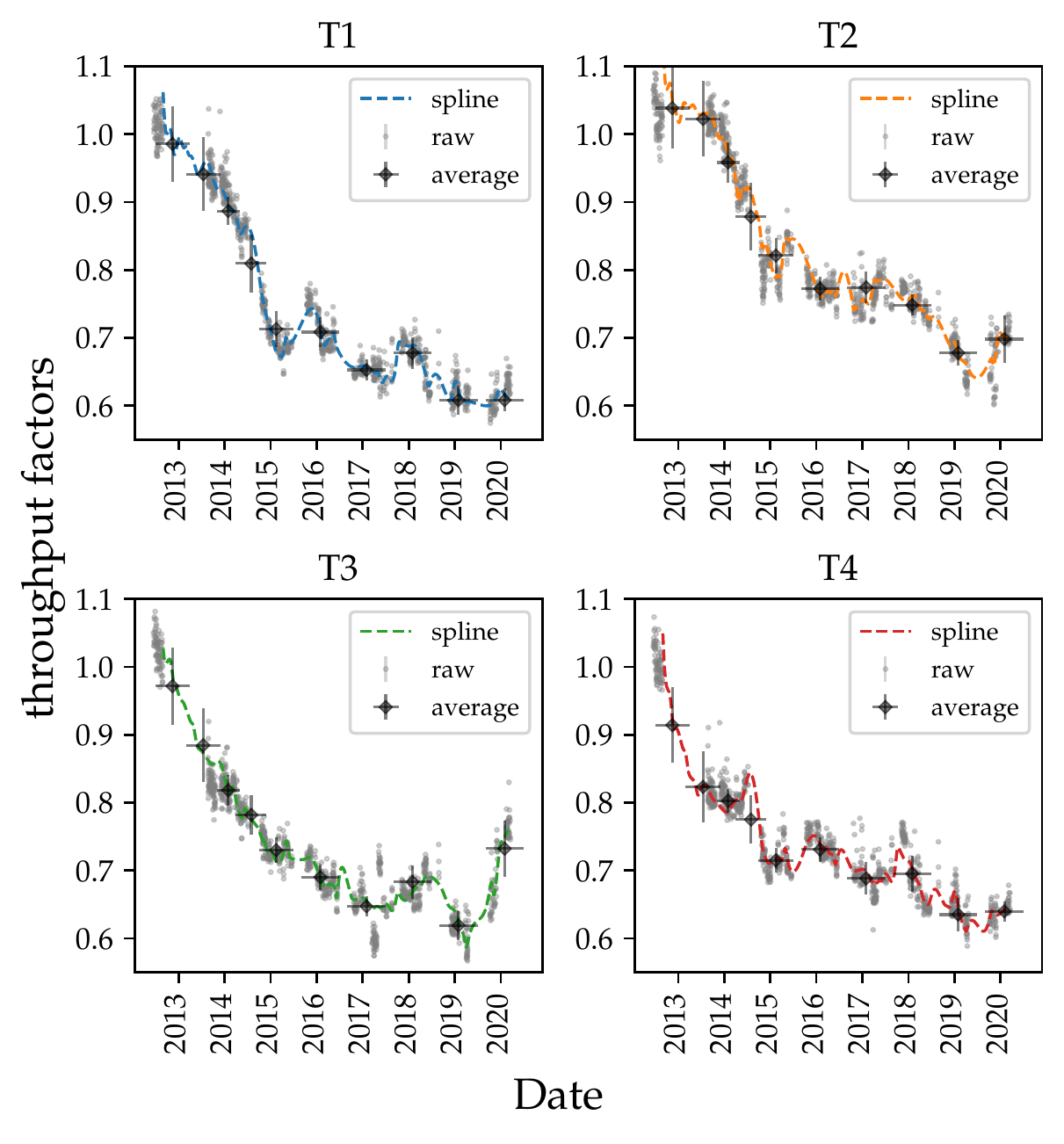}
\caption{Time-dependent change of the throughput factors calculated from photo-statistic gain factors and WDR reflectivity factors. Error bars show only statistical uncertainties.}
\label{fig:sfactors}
\end{figure}

The total throughput of each telescope, normalized to the reference values in the simulations, can be computed as the product of the gain factor $g_i$ (Figure \ref{fig:photostat_gfactors}) and the corresponding reflectivity factor $r_i$ (Figure \ref{fig:wdr_tfactors}). The resulting values and their evolution for each telescope are shown in Figure \ref{fig:sfactors} and table \ref{tab:irf_periods}. We refer to this total throughput factor as $t_i$ or simply throughput factor.

To ensure consistency, we computed the average throughput factors for each time bin with three different strategies: i) using the product of spline interpolations for gain factors and reflectivity factors and then computing the mean for each season, ii) using the closest reflectivity and gain measurement pairs to compute the total throughput and then estimating seasonal averages, and iii) using the average values of gain factors and reflectivity factors per instrument epoch and calculating their product. They all agree at the $\sim5\%$ level. 

The application of these throughput factors to the Cherenkov signals of the reference simulations allows us to produce IRFs that account for the throughput degradation of the VERITAS telescopes. The actual implementation and assessment of this calibration is discussed in the next section.

\subsection{IRF period definition}\label{sec:irfperiods}

VERITAS has undergone two major instrument upgrades. The first consisted on the relocation of T1 to its current position, improving significantly the trigger efficiency. The second upgrade targeted the camera, with improvements to the electronics and the PMT QE. This naturally generated three major instrument configurations,  
each with its own set of MC simulations, which provide an accurate model of the telescope performance. 

The current instrument configuration spans almost eight years of operations, during which the instrument has evolved significantly due to aging and our maintenance efforts. Consequently, we tune our IRFs and simulations in finer time bins so that each time bin or IRF period has throughput drops of at most $10\%$, comparable to the claimed systematic uncertainties on dish reflectivity and gains. Similarly, we require the throughput values at the limits of each bin to be covered by the statistical errors for that bin. Finally, the duration of each epoch is at most a single year, and approximately aligned with the observing cycle that begins with the end of the monsoon in September or October. 

With these criteria it is clear that for the last years, during which the evolution of the throughput was less dramatic, one IRF period per year is enough to provide the desired granularity and precision. For the first couple of years, however, the total throughput was rapidly evolving and it required finer time bins to fulfill our criteria. The resulting IRF periods are summarized in Table \ref{tab:irf_periods}.

\begin{table*}
\centering
\begin{threeparttable}
\caption{IRF periods for the current VERITAS instrument configuration. }
\label{tab:irf_periods}
\begin{tabular}{lllrrrr}
\toprule
 season     &   MJD range & data runs &  $t_1$            & $t_2$              & $t_3$             &  $t_4$ \\
\midrule
 2012-2013a & 56124-56367 & 63373-67410 & $0.986 \pm 0.056$ &  $1.039 \pm 0.059$ & $0.972 \pm 0.057$ & $0.914 \pm 0.056$ \\
 2012-2013b & 56367-56588 & 67411-70170 & $0.941 \pm 0.055$ &  $1.023 \pm 0.056$ & $0.885 \pm 0.054$ & $0.823 \pm 0.052$ \\
 2013-2014a & 56588-56778 & 70171-73235 & $0.887 \pm 0.021$ &  $0.958 \pm 0.030$ & $0.818 \pm 0.022$ & $0.803 \pm 0.018$ \\
 2013-2014b & 56778-56968 & 73236-75021 & $0.809 \pm 0.042$ &  $0.878 \pm 0.050$ & $0.782 \pm 0.029$ & $0.775 \pm 0.036$ \\
 2014-2015  & 56968-57242 & 75022-78239 & $0.713 \pm 0.027$ &  $0.821 \pm 0.026$ & $0.730 \pm 0.019$ & $0.715 \pm 0.019$ \\
 2015-2016  & 57242-57600 & 78240-82587 & $0.708 \pm 0.022$ &  $0.772 \pm 0.018$ & $0.690 \pm 0.018$ & $0.731 \pm 0.018$ \\
 2016-2017  & 57600-57966 & 82588-86848 & $0.653 \pm 0.016$ &  $0.774 \pm 0.023$ & $0.647 \pm 0.015$ & $0.688 \pm 0.024$ \\
 2017-2018  & 57966-58331 & 86849-90608 & $0.677 \pm 0.023$ &  $0.748 \pm 0.016$ & $0.683 \pm 0.025$ & $0.695 \pm 0.027$ \\
 2018-2019  & 58331-58696 & 90609-93829 & $0.608 \pm 0.021$ &  $0.678 \pm 0.019$ & $0.619 \pm 0.021$ & $0.635 \pm 0.024$ \\
 2019-2020  & 58696-59061 & 93830-96048 & $0.608 \pm 0.017$ &  $0.698 \pm 0.034$ & $0.732 \pm 0.042$ & $0.640 \pm 0.015$ \\
\bottomrule
\end{tabular}
\begin{tablenotes}
\small
{\raggedleft
\item {\em Note:} Definition of IRF periods for the current VERITAS instrument configuration with the covered dates, the corresponding data runs and the values of the total throughput corrections that need to be applied to the simulated event charges to obtain the calibrated MC and IRFs. Only statistical uncertainties were considered in this study.
}
\end{tablenotes}
\end{threeparttable}
\end{table*}

\section{Implementation of throughput calibration}\label{sec:throughput_corrections} 

This section discusses various strategies to implement the throughput calibration in IACTs, a description of the method adopted in the VERITAS software packages and the tests we performed to ensure the consistency of the results.

\subsection{Implementation methods}

MC simulations should accurately reproduce the response of VERITAS at any given time. 
We discussed in the previous section the impact of the time-dependent changes on the PMT gains and optical elements of the telescope. Reproducing all the details with the required accuracy while using a single MC production and a static set of IRFs is not an option if throughput is changing over time. Here, we concentrate on how we can implement the throughput calibration in our analysis framework to ensure consistent detector response over the entire operating life of the telescopes.

There are a number of strategies one could adopt to implement this calibration in the analysis, with varying degrees of accuracy that correlate with the computational cost and the parameter space. In the case of VERITAS, we need to cover the response of the instrument as a function of zenith angle of observation, azimuth angle, atmospheric profiles, telescope multiplicity, camera HV settings, wobble pointing offset, and NSB intensity.

Full MC productions that reproduce the changes in the throughput of the instrument over time and corrections to an existing MC production are both valid possibilities. Provided that the systematic errors on the calibration measurements are small, the former is potentially the most accurate and correctly reproduces the changes at the trigger level caused by throughput variations. However, it is computationally very expensive. It also requires producing a new complete MC model of the entire array and simulating $\gamma$-ray showers with the required statistics. Consequently we chose to explore corrections to the already existing MC simulations that were produced for the baseline instrument configuration. 

Calibration corrections could be applied on a run-wise basis (with typical duration of an observing run shorter than one hour) or on longer periods of time: from just one day to an entire year or even longer, during which the instrument response is measured to be approximately stable. Because of the monsoon shutdown in summer, which defines a natural observing season from about mid-September to mid-June, the fact that camera voltages (and hence gains) are adjusted at the beginning of each of these seasons and the slow cadence of WDR measurements, such seasons seemed a natural choice for VERITAS to define the time bins for the corrections. Exceptionally for the first two years, for which the throughput evolved rapidly, each season was divided into two shorter time bins. We refer to these time bins in which we apply the throughput corrections as IRF periods.

Throughput corrections can be introduced at different steps in the analysis of simulated showers and the production of the IRFs. A simple option, yet not fully accurate, is to calibrate simulated event images after they have been cleaned and the PMT traces have been integrated. The size of the event is modified using the changes in the total throughput of the instrument. Noise level changes can be either ignored or approximately corrected. Testing this option in VERITAS resulted in a reasonable reconstruction of event energies and $\gamma$-ray fluxes. However, the corrected simulated image shapes did not match the shape of the observed shower images. 

A natural improvement over this, and the solution adopted in this study, consists of correcting the signals of the simulated events before the PMT traces are integrated and the Cherenkov images are cleaned. Even though the trigger changes are still not properly modeled, noise levels change as fewer pixels containing only NSB are able to pass the image cleaning thresholds. 
Our tests resulted in reconstructed simulated showers which are directly comparable to observed data, making it easier to assess the accuracy of the corrections by comparing the distributions for the different image parameters. 
The main disadvantage is that the cleaning and Hillas parameterization of the simulated showers had to be repeated once per IRF period with each throughput factor $t_i$ on the signal integration, significantly increasing the computing time, storage requirements and analysis complexity needed to cover the entire parameter space. Producing IRFs for a single wobble offset with 10 time bins as described in this work, including two atmospheric profiles, and all zenith angles (8) and noise levels required (11), took several weeks of testing and computation time and required about $480\,\mathrm{GB}$ of storage for the final products. 

\subsection{Robustness of shower parameter reconstruction}

\begin{figure}[!htbp]
\centering\includegraphics[width=1\linewidth]{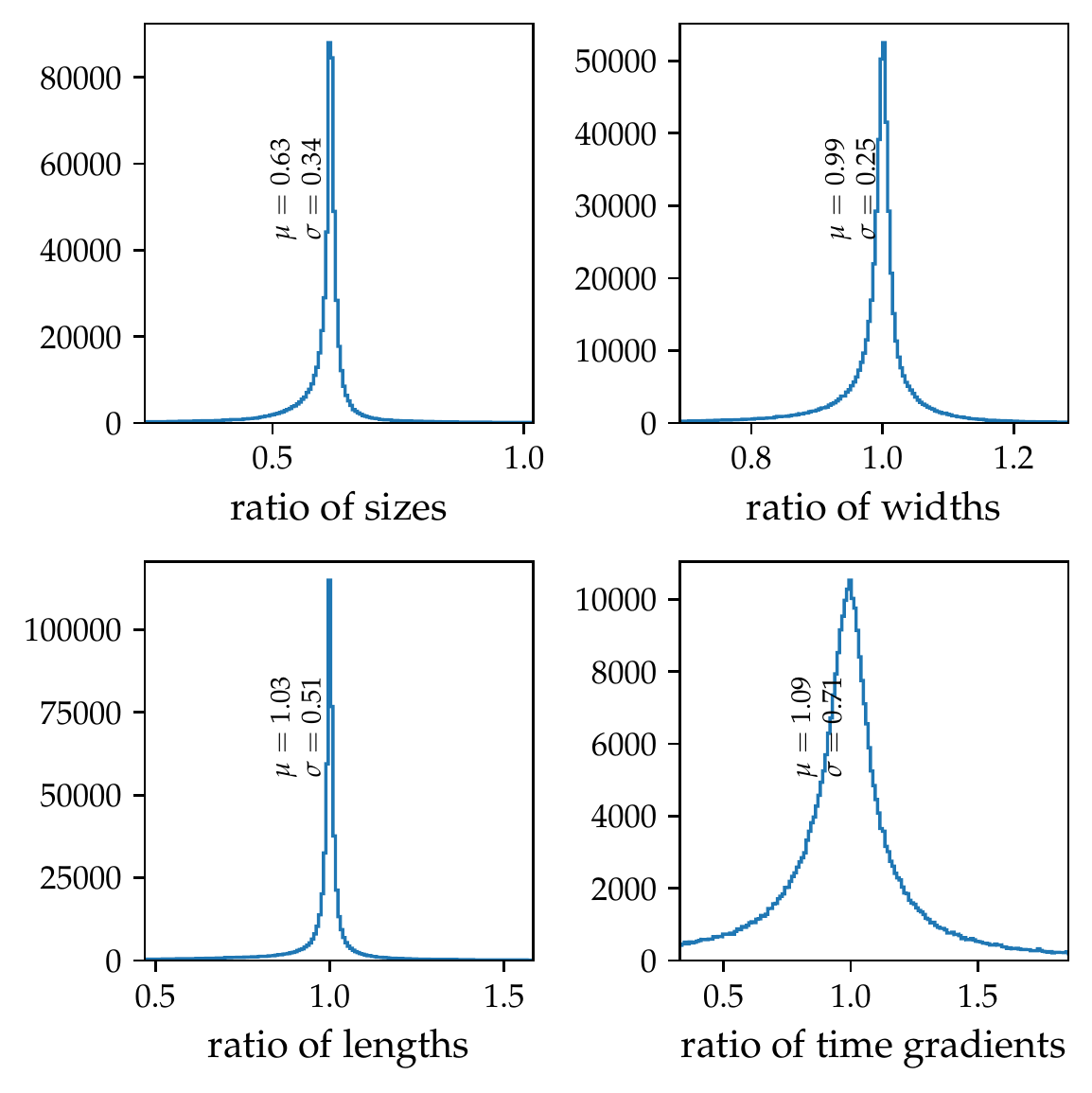}
\caption{Ratio between image parameters derived from MC simulations for the season 2019-2020 vs season 2012-2013a (Telescope T1 only, winter simulations, noise level of $100\,\mathrm{MHz}$, and no size cuts). The vertical axis of each panel shows the number of events within a given ratio bin.}
\label{fig:ratio_mcvariables_v6}
\end{figure}

This work is based on the assumption that we can correct the simulated PMT traces with just two global factors, one due to the camera gain $g_i$ and the other due to the dish reflectivity $r_i$. In principle, this should not distort the shape of the shower images; instead we expect it to alter the size of the cleaned events. 
To check if this is the case, we extracted the same simulated events corresponding to two different IRF periods (2012-2013a and 2019-2020) and calculated the ratio of values for some of their key geometrical parameters (size, width, length, and time gradient along the major axis) between the two periods. Figure \ref{fig:ratio_mcvariables_v6} shows histograms of the obtained ratios together with the mean and the standard deviation of the obtained distribution. It can be seen that only the size of the events is significantly shifted ($\mu \simeq 0.63$, $\sigma  \simeq 0.34$) when we scale the simulated pixel traces. As this is a statistical comparison between individual simulated showers obtained with independently generated IRFs, the large width of the distributions for some of these ratios is not a concern for this study.

\subsection{Impact of throughput changes on VERITAS performance}\label{sec:throughput_effects} 

The evolution of the total throughput can affect the reconstructed fluxes in two different ways. The relation between the size of an event and its energy changes as discussed in section \ref{sec:recenergies}. Additionally, some events are no longer reconstructed, as their integrated charges are not large enough to pass the hardware trigger, analysis, and reconstruction cuts. This has a significant impact for the lowest energies, as is shown in the effective areas of the telescopes, discussed in section \ref{sec:effareas}.

\subsubsection{Effect on reconstructed energies}\label{sec:recenergies}

\begin{figure}[!htbp]
\centering\includegraphics[width=1\linewidth]{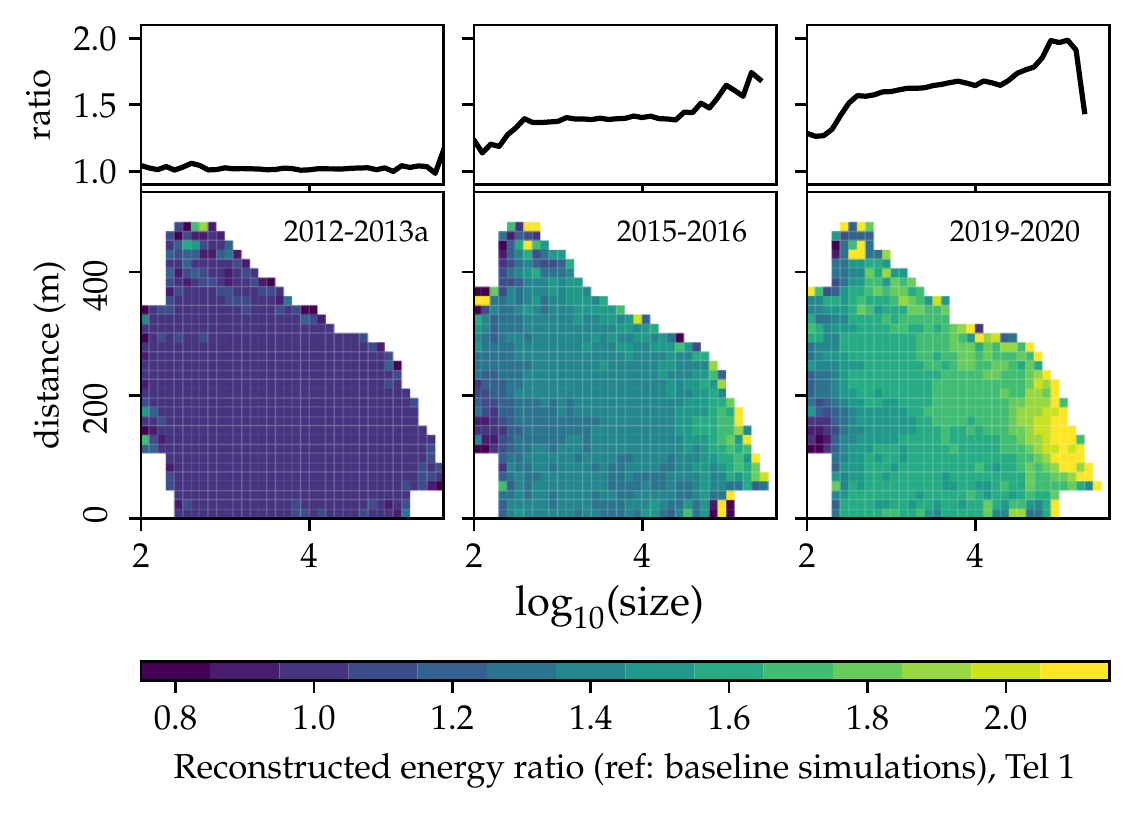}
\caption{Relative changes in energy reconstruction induced by the evolution of the throughput of the instrument. The ratio of throughput-calibrated energies with respect to the uncorrected energies is presented in two formats: As 2D lookup tables, with the ratio as function of impact distance and size of the simulated events ({\bf bottom panels}); curves with the distance-averaged lookup-tables, showing the ratio as a function only of size ({\bf top panels}).}
\label{fig:energytable}
\end{figure}

One of the main effects of the application of correction factors to the measured digital traces is the improvement of the energy reconstruction. Lowered gain and reflectivity naturally result in smaller images and consequently lower estimated energy if the IRFs do not account for throughput variations. 
The energy reconstruction method applied uses lookup tables, similar to the ones described in section \ref{sec:gamma_hadron_sep}. They represent the energy as a function of size and distance to the telescope in a coordinate system perpendicular to the arrival direction of the air shower event, the NSB, the wobble offset of the observation and the atmospheric profile. These tables are filled using simulated $\gamma$-ray events by calculating the median and the standard deviation of the distribution of true energies ($\mathrm{E_{MC}}$) for each bin in the grid. 

Figure \ref{fig:energytable} shows the ratio between the reconstructed energies of the events whose traces have been corrected to account for the drop in telescope throughput and the uncorrected case. That ratio is represented, following the format of the lookup tables, as a function of impact distance and size of the simulated event for a fixed set of values for the other parameters.

For a given noise level and zenith angle, the size of the event scales almost linearly with its energy. The exception would be the highest energies. In that regime, the accumulated charge may hit a maximum value due to FADC saturation, affecting the linear relation between size and energy previously described. 
Since the events migrate to lower sizes for a fixed energy as the telescope throughput drops, one can easily see the effect of its evolution with time once we calibrate the throughput in the analysis: for a given image size and distance of the event, the reconstructed energy is higher in recent years than it was right after the upgrade of VERITAS (years 2012-2013).

\subsubsection{Effect on flux reconstruction and effective  areas}\label{sec:effareas}

\begin{figure}[!htbp]
\centering\includegraphics[width=1\linewidth]{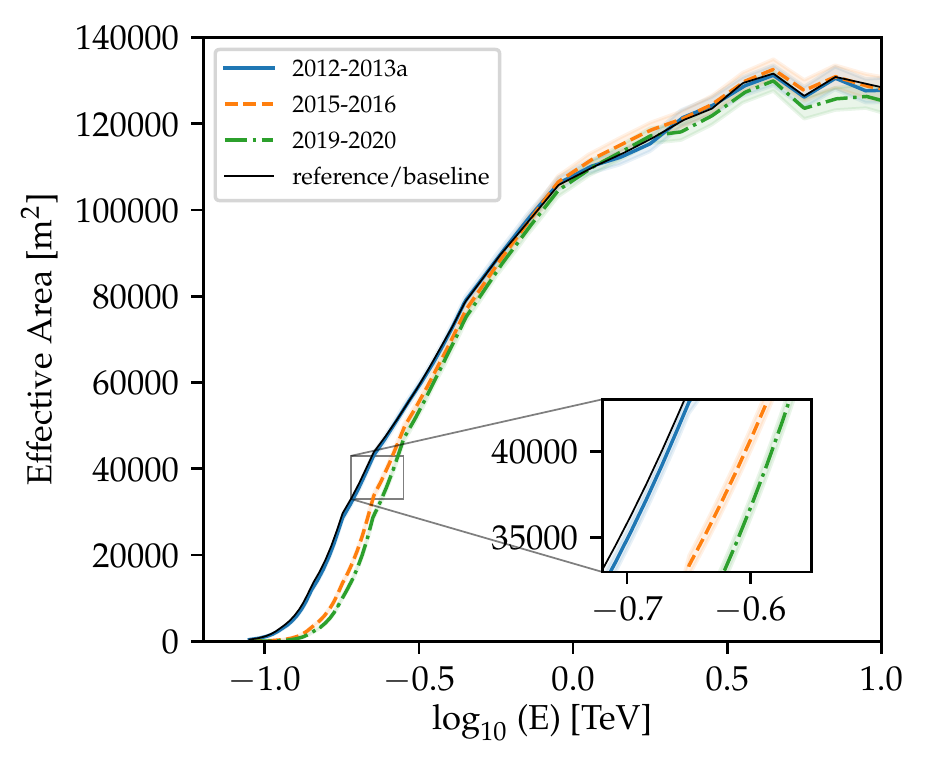}
\caption{Evolution of the effective area for different instrument epochs, for a zenith angle of observation of $20^\circ$, noise level of $100\,\mathrm{MHz}$, moderate cuts and a minimum of two images per event.}
\label{fig:effareacomparison}
\end{figure}

The impact of the total throughput on the effective area is shown in Figure \ref{fig:effareacomparison}. In that figure, the effective area is computed as a function of true energy in each period for a fixed position in the parameter space (zenith angle of $20^\circ$, noise level of $100\, \mathrm{MHz}$, moderate cuts, a minimum of two images per event, and $0.5^\circ$ wobble offset).

There is insufficient background data to optimize the cuts individually for every period. The reason is that we require data that were taken under very good weather conditions, fields without strong $\gamma$-ray sources, and dividing the remaining data sample into zenith angle bins. Consequently, we used the same set of cuts optimized using data from the entire post-upgrade array configuration. This artificially enhances the differences seen at low energies in Figure \ref{fig:effareacomparison}, since fewer events have the size required to pass the analysis cuts in the latest periods. Reducing the size cut over time to palliate this effect is not an option as it would result in the inclusion of worse quality events in the analysis. Above $\sim 300\,\mathrm{GeV}$, away from the energy threshold of the analysis, almost no events are lost despite the decreased optical throughput of the telescope. 

\subsubsection{Effect on the energy threshold of the analysis}

Due to the statistical nature of the reconstruction methods used for IACTs, the energy threshold of the analysis is defined as the energy at which the effective collection area is  $10\%$ of the maximum effective area reached with the telescope at any energy. This definition ensures that a significant fraction of the events have passed the cuts and that the individual shower images and traces are comparatively less affected by noise artifacts. It does not imply that lower energy particles are undetectable, but the percentage of particle showers below this energy quickly drops as the energy decreases. The detected emission at energies lower than the energy threshold may still be significant for sources with steep spectra. 

\begin{figure}[!htbp]
\centering\includegraphics[width=0.8\linewidth]{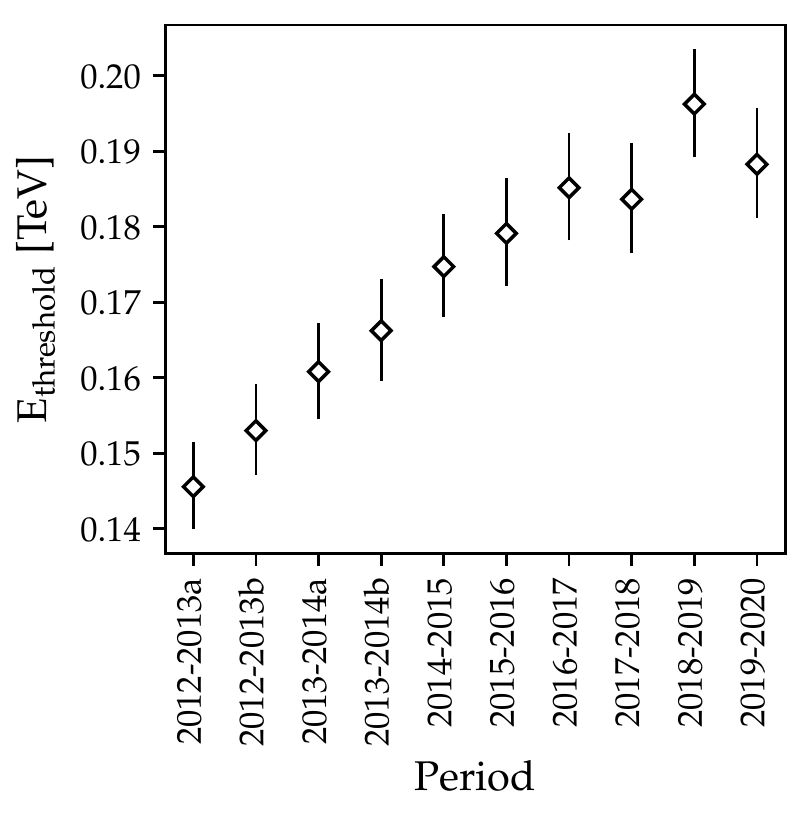}
\caption{Energy threshold of the analysis estimated from simulated events and using moderate cuts, zenith angle of $20^\circ$, and noise level of $100\,\mathrm{MHz}$. It is defined as the energy for which the effective area becomes $10\%$ of the maximum effective area.}
\label{fig:energy_threshold}
\end{figure}

Figure \ref{fig:energy_threshold} shows the evolution of the energy threshold of the  analysis with VERITAS. Uncertainties of the energy thresholds shown in the figure are calculated accounting for two independent contributions: 
\begin{enumerate*}[label=\roman*)]
\item the energy bin width of the effective area histogram,
\item the uncertainty on the effective area at the center of the bin.
\end{enumerate*}

\subsubsection{Effect on the differential sensitivity}\label{sec:diffsens}

\begin{figure}[!htbp]
\centering\includegraphics[width=0.8\linewidth]{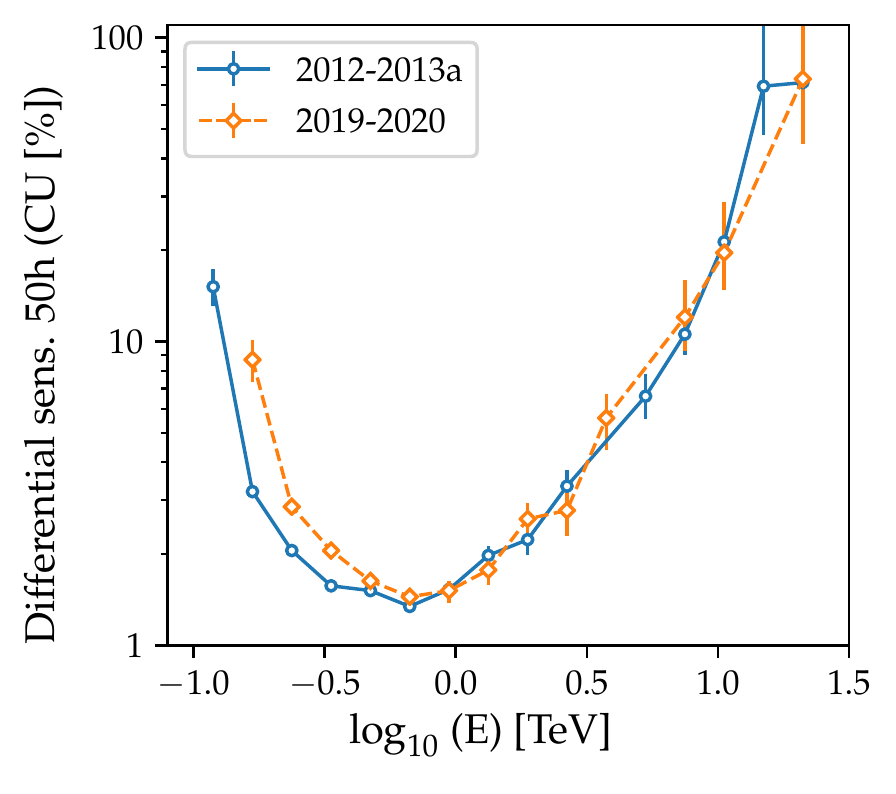}
\caption{Differential sensitivity with $50\,\mathrm{h}$ of VERITAS observations calculated using different low zenith angle Crab Nebula data, with winter atmospheric profile conditions. The same cuts as in Figure \ref{fig:effareacomparison} are used. The sensitivity is defined in terms of the flux of the Crab Nebula (in Crab units, CU) assuming it is a reference astrophysical object} at VHE energies.
\label{fig:diffsensitivity}
\end{figure}

We can define the sensitivity of a $\gamma$-ray instrument as the minimum flux that such an instrument can detect for a given amount of exposure. 
Simulating a realistic sample of background data is challenging, so it is common practice to use the Crab Nebula data as a reference object and to use its flux as the unit system for sensitivity studies. 
In the following, the differential sensitivity is computed as the minimum photon flux required to achieve a 5$\,\sigma$ detection in $50\,\mathrm{h}$ of observation time (using the statistics described in \cite{1983ApJ...272..317L}) in each energy bin, assuming five equal-width bins per decade on the logarithmic energy axis. Additionally, at least ten signal events are required per energy bin.

The differential sensitivity is presented in Figure \ref{fig:diffsensitivity}, based on good quality Crab Nebula data, which were collected under dark sky conditions, and at low zenith angle to minimize the atmosphere's influence. We show the results for BDT-based moderate cuts with a minimum telescope multiplicity of two, which provides a balance between optimizing sensitivity and maintaining a low energy threshold. Results employing softer cuts and harder cuts were obtained and compared between seasons, yielding similar conclusions. 

It can be concluded that throughput degradation affects mostly the lowest energies,  where some dim shower events are lost or mistakenly classified as hadronic showers. At high energies, the performance is similar between the different periods, unaffected by the changes in the throughput. 

\section{Validation of the throughput calibration on real data\label{sec:validation_realdata}}

\subsection{Comparison between real data and MC simulations}

The validity of the estimated total throughput correction can be tested by comparing image and stereo parameter distributions obtained with data and throughput-calibrated simulated events for every IRF period.
If MC showers are not calibrated with the correct throughput, the shower shapes will be distorted and the estimated shower parameter distributions will not match those derived from data.

In this work, we compared MSCW and MSCL distributions of simulated and observed data for six reconstructed energy bins: $\log_{10} E_{rec} \in$ $[-1.0,-0.7]$, $[-0.7,-0.3]$, $[-0.3,0.0]$, $[0.0,0.3]$, $[0.3,0.7]$, $[0.7,1.0]$ (energy measured in TeV), and for all the considered IRF periods. 

The results, with the last two energy bins combined to increase statistics, can be seen in Figure \ref{fig:mscw_data_mc}. More detailed comparisons are shown in appendix \ref{sec:appendix_mcdatacomparison}. In all cases, data and simulated events have similar distributions for both parameters, almost symmetric, with similar widths and centered at zero (as expected for $\gamma$-ray events). No significant evolution was observed in these parameters with time. The only exception is the highest energy bin, which has a wider distribution of MSCW values for real shower events. This is caused by the high energy events, $\sim 10\,\mathrm{TeV}$, being subject to difficulties in calibrating of the low-gain readout channel, saturation effects, and leakage of many shower images outside of the camera. Moreover, its scientific impact is limited because statistical uncertainties almost always dominate at $\sim 10\,\mathrm{TeV}$ due to the small photon fluxes emitted at these energies from astrophysical sources.

\begin{figure}[!htbp]
\centering\includegraphics[width=\linewidth]{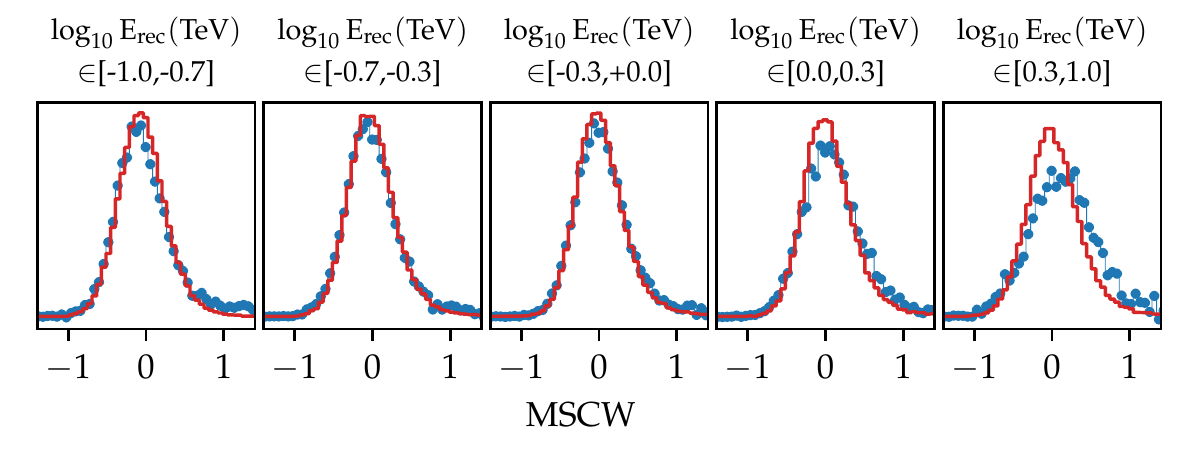}
\caption{Average distributions of MSCW values, in different energy bins, for the data (blue points) and simulation (red  curve) in the entire post-upgrade period. They were both generated after throughput calibration. Data events were obtained from a sample of Crab Nebula observations. 
Due to the small total number of events, the last two energy bins are combined into a single bin with energies $\mathrm{\log_{10} E_{rec}} \mathrm{(TeV)}  \in [0.3,1.0]$.}
\label{fig:mscw_data_mc}
\end{figure}

\subsection{Inter-telescope calibration}\label{sec:inttelcal}

Each telescope that detects a $\gamma$-ray-induced shower provides an independent estimation of the primary particle energy. This enables us to test how well the throughput of each telescope is calibrated against that of the other telescopes.

\begin{figure}[!htbp]
\centering\includegraphics[width=1\linewidth]{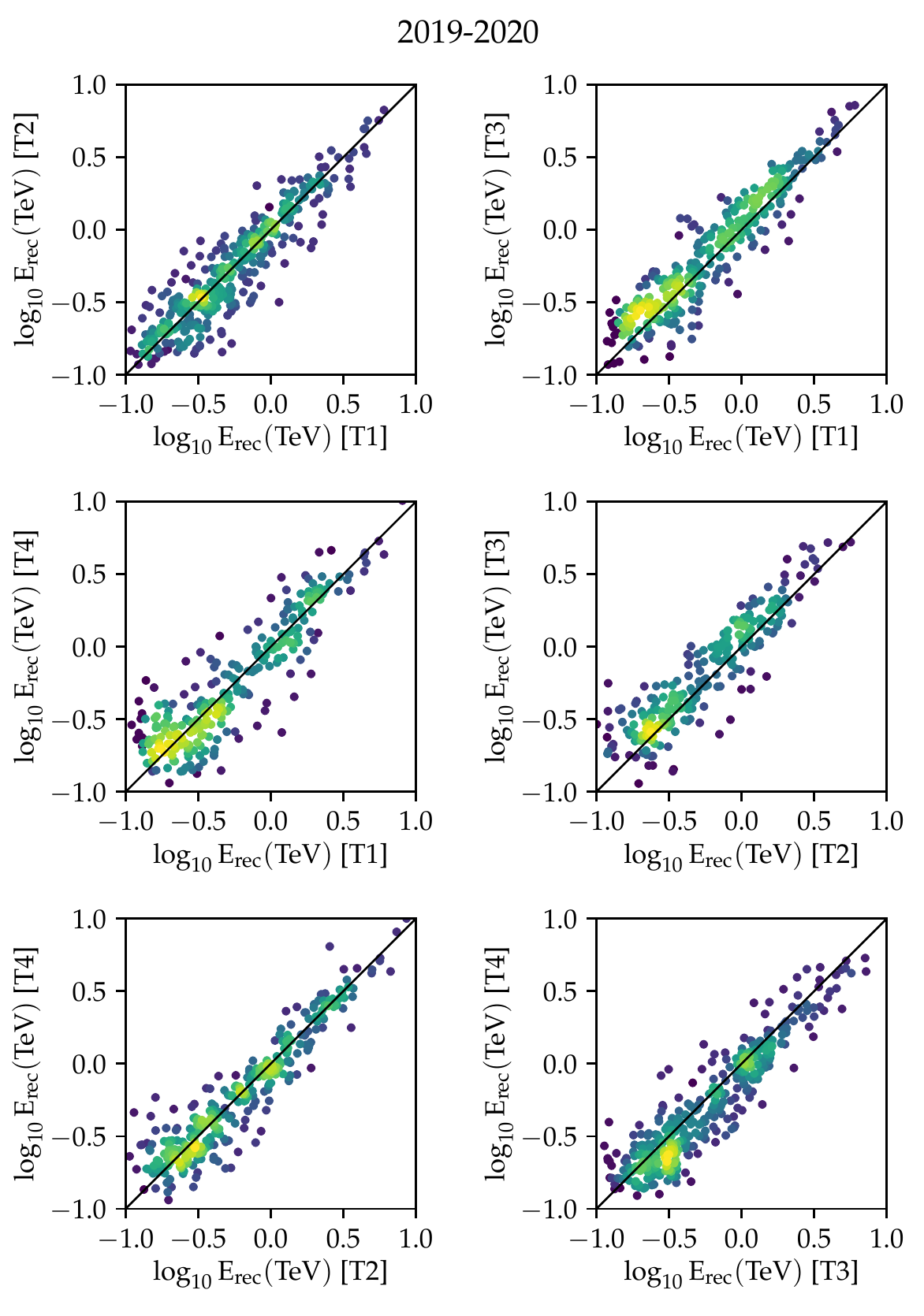}
\caption{Inter-telescope calibration: reconstructed energies for $\gamma$-ray-like events from Crab Nebula observations, with a similar (difference less than $20\%$) distance from each telescope to the reconstructed shower core. The density of events in the parameter space is color-coded, with yellow values representing more events and blue values representing fewer events. A 1:1 relationship for the event energies is plotted in black, corresponding to perfect inter-telescope calibration. For compactness, we only show the results for 2019-2020.}
\label{fig:sizeintertel}
\end{figure}

There are different ways of performing such a test \citep{2003APh....20....1H, 2003APh....19..221L, 2016APh....75....1M}. One possibility involves selecting, for each pair of telescopes, events with a similar distance to each telescope $R$ (in our case, within $\pm20\%$). Such events should have approximately the same reconstructed energy if the optical and camera calibration of the telescopes is appropriate. Alternatively, one can compare the energy reconstructed by a given telescope, $\mathrm{E}$, against the average energy reconstructed by the array, $\mathrm{E_{mean}}$. Figure \ref{fig:sizeintertel} illustrates the first approach (telescope pairs compared against each other) for 2019-2020. 

\begin{figure}[!htbp]
\centering\includegraphics[width=0.9\linewidth]{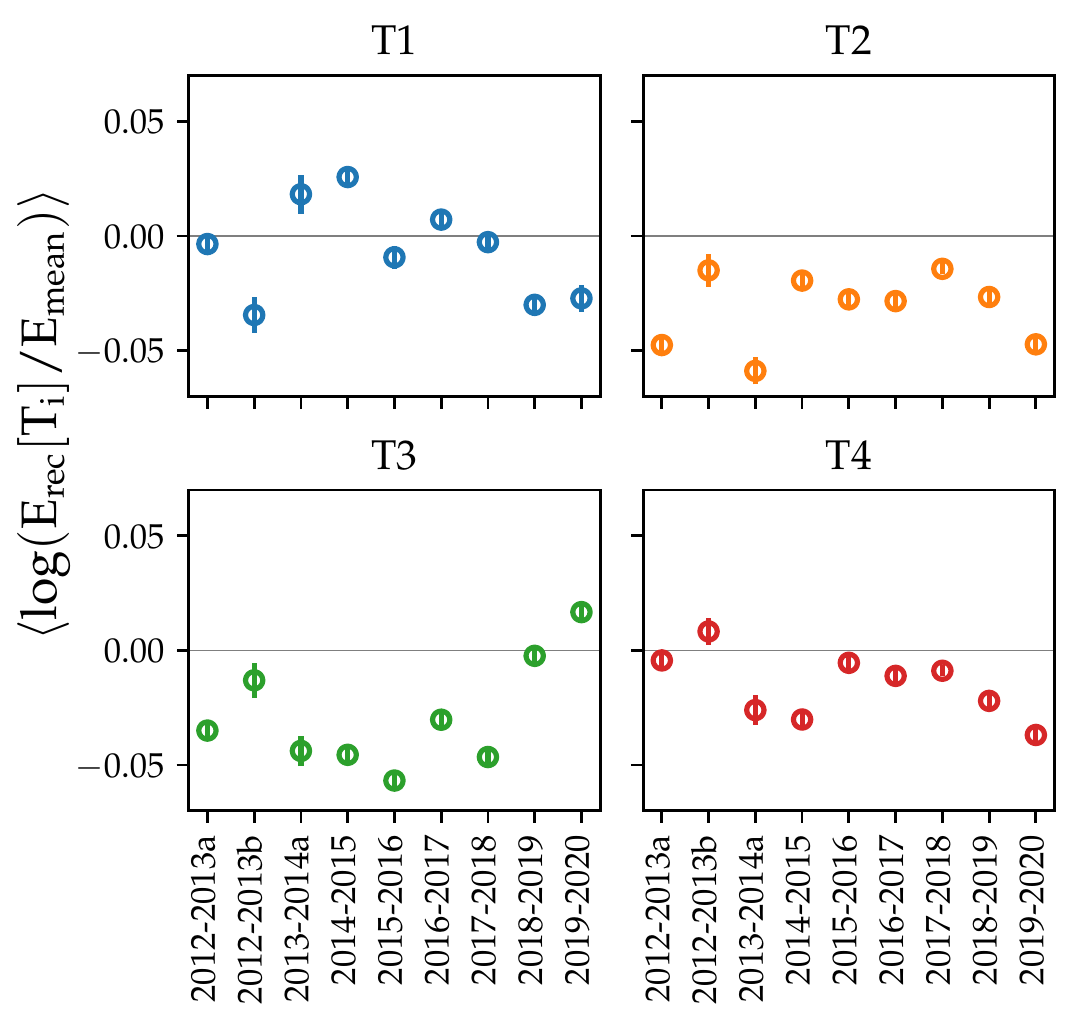}
\caption{Mean of the $\log_{10} \mathrm{(E_{rec} [T_i] / E_{mean}})$ values for selected $\gamma$-ray-like events and its standard deviation, for each telescope and season, where $\mathrm{E_{rec} [T_i]}$ is the reconstructed energy of telescope $i$ and $\mathrm{E_{mean}}$ the average reconstructed energy from the entire array, using the same weights for all telescopes. For a given epoch, the energies reconstructed by the different telescopes %are compatible within uncertainties, while the absolute value does 
do not deviate on average more than $\sim 10\%$ with respect to the mean energy reconstructed with the entire array. \label{fig:energy_dev_pertel}}
\end{figure}

We repeated these measurements for all IRF periods. For each period, we calculated the mean $\log_{10} \mathrm{(E_{i})} - \log_{10} \mathrm{(E_{mean})}$ and its standard deviation, shown in Figure \ref{fig:energy_dev_pertel}. The reconstructed energy of real $\gamma$-ray like events deviate, on average for each telescope and season, no more than $\sim 10\%$ with respect to the mean energy reconstructed by the entire array. If the throughput calibration had not been implemented successfully, we would see larger differences in the energy measured by the individual telescopes. Telescopes with faster throughput degradation would provide smaller reconstructed energy estimates.

\subsection{Reconstruction of the Crab Nebula flux}

The reconstructed flux and spectral shape for any measurement depends on the flux calibration of the instrument. The Crab Nebula is the closest object to a standard reference object for VHE observations as it is one of the brightest sources known that has a stable emission. Therefore, we used observations of this source for the last tests to validate both the throughput measurements and the actual implementation of the throughput calibration. 
This study was performed using the same Crab Nebula dataset and the same cuts as used in Section \ref{sec:diffsens}. Two figures summarizing the results of this study are obtained:

\begin{figure*}[!htbp]
\centering\includegraphics[width=0.85\linewidth]{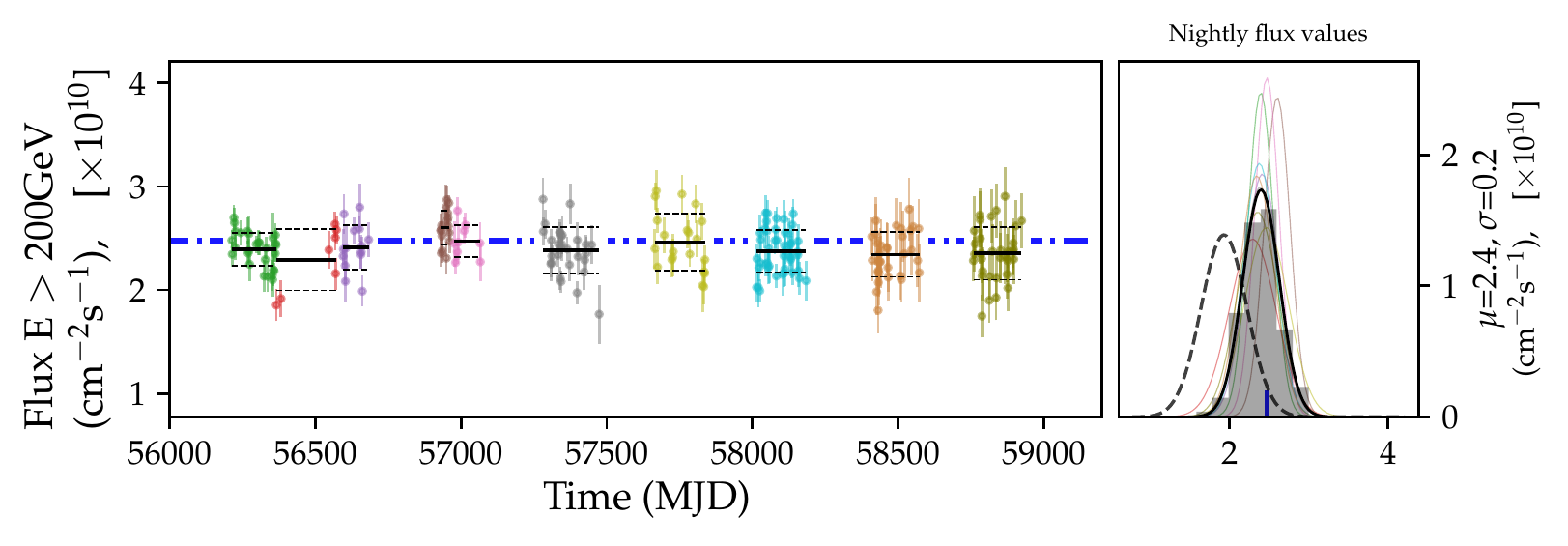}
\caption{Reconstructed flux of the Crab Nebula above $200\,\mathrm{GeV}$, using 1-day bins and IRFs that correctly match the instrument throughput for each period. Results are obtained applying moderate cuts to runs with mean elevation $>50^\circ$. {\bf Left panel:} Each color represents an IRF period. The blue dot-dashed curve represents the reference Crab Nebula spectrum of \cite{2015ICRC...34..792M} integrated above $200\,\mathrm{GeV}$, horizontal solid black lines represent the season-average fluxes, while dashed horizontal curves show the standard deviation of the fluxes for each season. {\bf Right panel:} Shown in gray is the distribution of fluxes for all seasons combined, with a fit to a Gaussian shape as a solid black line. The dashed black curve shows the equivalent result when throughput changes are not taken into account in the IRFs. The vertical blue dashed line shows the reference Crab Nebula flux from \cite{2015ICRC...34..792M}.}  
\label{fig:light_curves_crab}
\end{figure*}

\begin{figure}[!htbp]
\centering\includegraphics[width=0.8\linewidth]{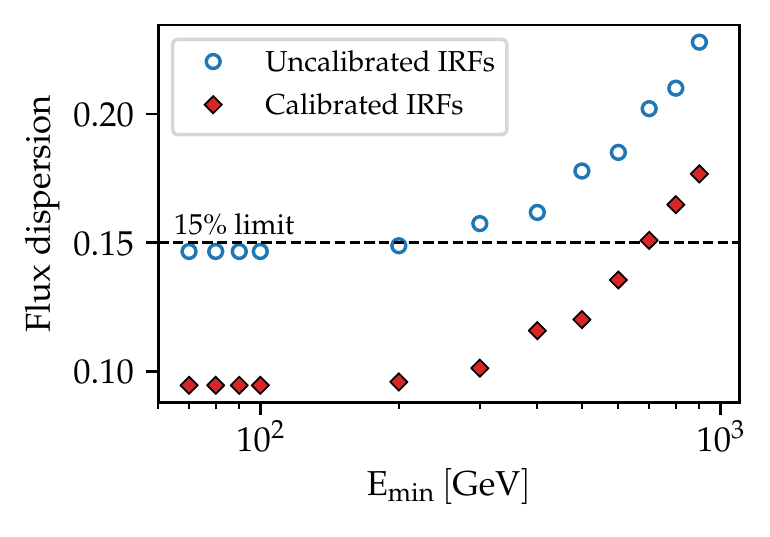}
\caption{Flux dispersion ($\sigma/\mu$) as a function of threshold energy for light curves similar to the one of Figure \ref{fig:light_curves_crab}, using moderate cuts. Taking into account the throughput evolution brings down the statistical uncertainties from $\sim 15\%$ baseline to $\sim 10\%$.}
\label{fig:lcenergy_fluxdispersion}
\end{figure}

\begin{figure*}[!htbp]
\centering\includegraphics[width=0.8\linewidth]{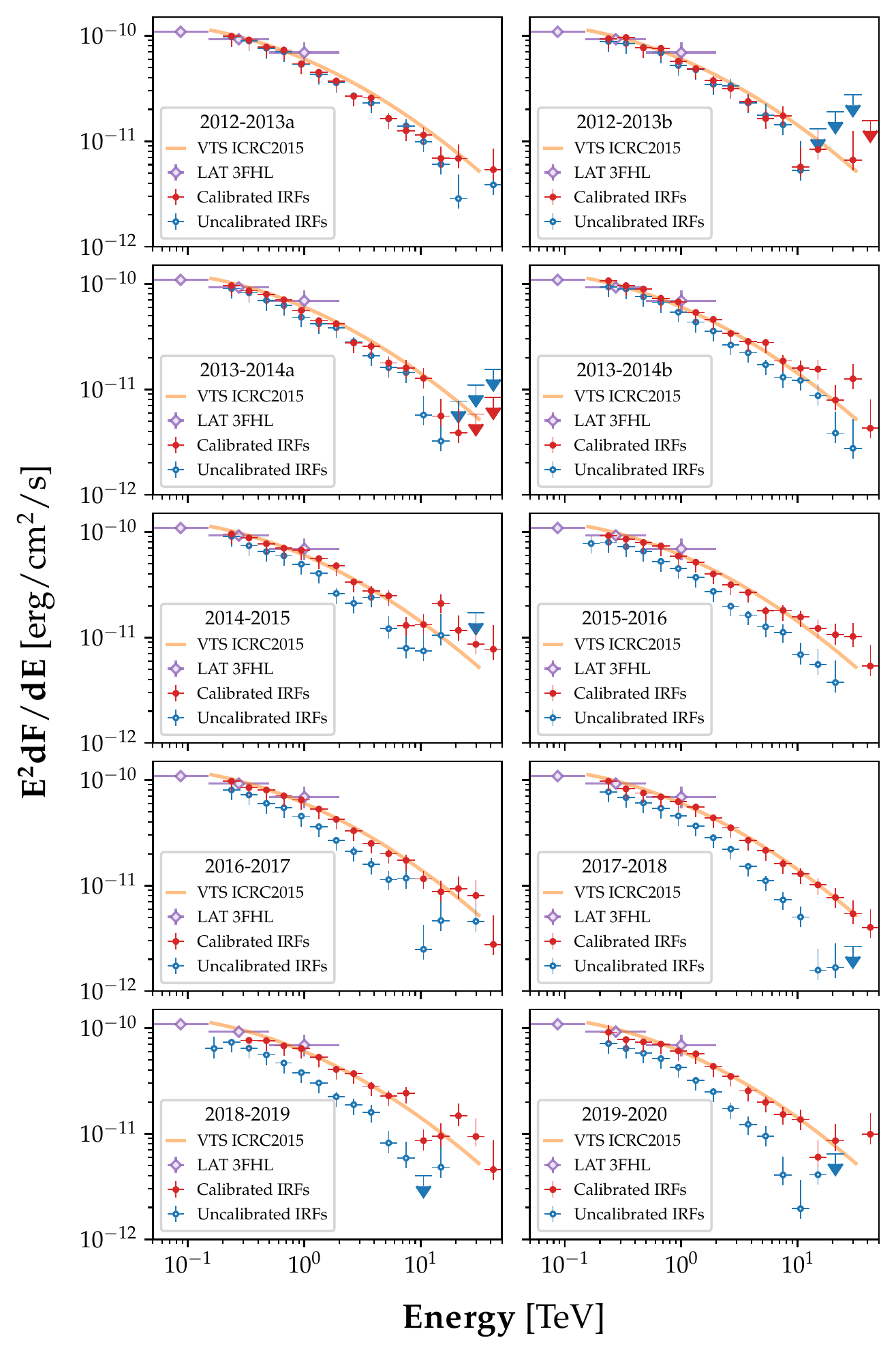}
\caption{Collection of spectral energy distributions obtained with moderate cuts for each IRF period. Purple open diamonds represent measurements of the Crab Nebula spectrum from the 3FHL \citep{2017ApJS..232...18A}, the orange curve shows the Crab Nebula spectrum published by VERITAS in \cite{2015ICRC...34..792M}, open blue points show the fluxes that would be obtained if throughput is not calibrated during the reconstruction of $\gamma$-ray showers and instead the MC model from 2012 is employed. Finally, red solid points show the spectral points that are obtained once the pulse signals are calibrated to generate correct IRFs for each period.}
\label{fig:energy_spectra_crab}
\end{figure*}

\begin{itemize}
\item Light curves with daily binned fluxes above $200\,\mathrm{GeV}$ for BDT-based moderate cuts (see Figure \ref{fig:light_curves_crab}). Datasets corresponding to the different IRF sets are plotted with different colors. 
Flux points versus time are shown in the left panel and a histogram of the flux points is presented in the right panel. As can be seen from the figure, every season shows an average flux compatible within $\pm 1\,\mathrm{\sigma}$ with the flux of the Crab Nebula reported in \cite{2015ICRC...34..792M}. In addition, 
the spread of the flux points (standard deviation $\sigma$ over the mean $\mu$ of the distribution) shown in the right panels remains on the level of $\lesssim 10\%$ for the selected cuts. 
This value is significantly lower than the one obtained if no throughput calibration is applied. In that case, the spread would be $\gtrsim 15\%$. 
The relationship between the energy threshold of the light curve and flux dispersion is shown in Figure \ref{fig:lcenergy_fluxdispersion}.

\item Energy spectra for the different periods, using the same cuts (Figure \ref{fig:energy_spectra_crab}), showing simultaneously for each period the energy spectrum obtained using the uncorrected IRFs (blue open circles), the energy spectral points for the corrected IRFs (filled red circles) and, for comparison purposes, the published spectrum from the {\em Fermi}-LAT 3FHL catalog \cite{2017ApJS..232...18A} and the Crab Nebula spectrum published by the VERITAS collaboration in \cite{2015ICRC...34..792M} (orange curved line), which is based mostly on data from seasons that were minimally affected by the throughput degradation. 
\end{itemize}

\section{Systematic uncertainties}\label{sec:systematic_uncertainties}

The estimation of systematic uncertainties affecting the reconstructed $\gamma$-ray energies and fluxes covered in this section is based on calibration measurements, MC simulations, and reasonable assumptions. Overall, we could identify the following components of the systematic uncertainty budget impacting the throughput calibration and the determination of precise IRFs:

\begin{itemize} 

\item Accuracy and stability (time and temperature dependence) of the photo-electron response, including the precision of the pixel-to-pixel relative gain correction. 
Absolute pixel gains are corrected through high-voltage adjustments (``flat-fielding'') every year at the beginning of the VERITAS observing season, in September or October.
This contributes to reduce the photo-electron response drift of the PMTs over long periods of time. Relative gains likely have a small contribution to the total systematic uncertainty if the appropriate Flasher runs are used in the analysis.

\item QE of the PMTs and collection efficiency at the first dynode. After several years of operations, a comparison of used and unused PMTs shows no significant evolution of these parameters over time. 

\item Mirror reflectivity and its wavelength dependence after correction for degradation.

\item The optical PSF, determined by mirror alignment accuracy and checked once a month. Mirror facets are re-aligned after detection of significant deviation from the expected position.

\end{itemize}

Independently of the throughput calibration, the overall systematic uncertainty budget of the VERITAS instrument is affected by:

\begin{itemize}

\item The effect of broken pixels and electronic channels, which are not modeled in the MC simulations. The fraction of unavailable pixels per camera is typically less than 5\%.

\item The Winston cone efficiency, not covered in our study but included in the simulations, with an estimated contribution to the optical throughput uncertainty of $\sim 4\%$.

\item The treatment of shadowing elements on the telescopes in the MC simulations, including the camera housing and support structure. Smaller elements like cross bars are not included in the simulation model. The impact of these simplifications introduces a systematic uncertainty of less than 1\%.

\item The systematic limit of the VERITAS pointing monitor, which is less than 20 arc seconds, much smaller than the optical PSF of the telescopes. Its contribution to the systematic uncertainty on the photon flux is therefore <3\%.

\item The ability to calibrate the linearity of the readout and the transition from high to low gain using flasher pulses at eight (fifteen starting from 2016) different levels of brightness. 
The pulse shape of the low-gain channel is roughly twice as broad than that for high gain and exhibits a complex broadening for extremely bright signals from nearby showers with energies of tens of TeV. 
This complex behavior is not accurately reproduced in the current MC simulations, leading to an increase of the systematic uncertainties for the highest energies.
The throughput changes discussed in this paper lead additionally to a change of pulse brightness required to activate the low-gain chain. 
Our correction method, which cannot account for this hardware effect, leads to a mismatch in the activation point between the simulations and the real data. 

\item The analysis and reconstruction pipelines, Eventdisplay and VEGAS. While the code development is independent, the analysis and reconstruction techniques employed in both pipelines are similar and they use the same MC simulations to produce the IRFs. Based on a long-term comparison of scientific results produced with both tools, we estimate a $10\%$ systematic uncertainty due to the analysis pipeline used.

\end{itemize}

\begin{table}[!htbp]
\centering
\caption{\label{tab:systematics} Summary of systematic uncertainty estimations.}
\begin{tabular}{@{}lr@{}}
\toprule
Type & Flux  \\
\midrule
Photo-electron response and gain & 5\% \\
Signal pulse shape & 5-10\%\\
Low-gain modeling & 5-10\% \\
Efficiency of photo detection & $<5$\%\\
Unavailable pixels & 3-5\%\\
Shadowing & 1\%\\
Mirror reflectivity & 10\%\\
Winston cone efficiency & 4\% \\
Point spread function & 5-10\%\\
Atmospheric profiles, & \\
Absorption, scattering & 10-15\% \\
Analysis & 10\%\\
\midrule
Total & $\sim 25\%$ \\
\bottomrule
\end{tabular}
\end{table}

In general, the total systematic uncertainties are dominated by uncertainties on telescope throughput, covered by this work, and the approximations in the modeling of the atmosphere above the observatory.

As for the latter, the development of extensive air showers is determined by the vertical density, temperature, and humidity profiles. The intensity of Cherenkov light is influenced by the amount of absorption and scattering on atmospheric molecules or aerosols. 
While VERITAS does not operate during the warmest summer months, and two seasonal atmospheric profiles are used to correct for the large atmospheric changes, the systematic error due to inaccurate atmospheric models is approximately $10-15\,\%$. 
This includes day-to-day variability of the atmosphere, which is monitored by a combination of a weather station, a commercial ceilometer, and three infrared pyrometers. To limit its impact on the total systematic errors, observing periods of inferior quality (e.g., due to clouds) are flagged and removed from the analysis for most publications.

The total systematic uncertainty is estimated by summing quadratically the individual components listed in Table \ref{tab:systematics}, ignoring possible correlations.
As a caveat, some of the mentioned sources of systematic uncertainty are energy dependent and introduce an error on the reconstructed spectral slope, resulting in larger uncertainties for the steepest spectra. 
The total systematic uncertainty on the absolute flux level is estimated to be $\sim 25\%$ for a $\gamma$-ray source with a spectral index of $\approx 2.5$. Starting from $\sim 10\mathrm{\,TeV}$, where the calibration of the low- and high-gain channels and saturation effects begin to be important, an additional $5-10\%$ would have to be added in quadrature. 
Similarly, we estimate the systematic uncertainty on the spectral index to be $\pm 0.2$ for sources with Crab Nebula-like spectra. For sources with steeper spectra, the corresponding photon flux is dominated by the emission at the lowest energies, where many of the systematic uncertainty components that we mentioned become most relevant. In addition, the impact of the energy scale errors on the absolute flux become significantly larger and their estimation requires a case-by-case study, beyond the scope of this document. 

Additional sources of systematic uncertainties not directly linked to the telescope performance have not been included in this work. Nonetheless, some might have relevant implications in long-lived ground-based astronomical installations, including VERITAS. A good example of this is the long-term evolution of NSB \citep{2000PASP..112..566M} due to an increased human activity over time and changes in street lamp technology: from sodium lamps to LED-based illumination, likely brighter at short wavelengths \cite{2017MNRAS.467.2966S}, where IACTs are most sensitive. Such long-term variations are likely more evident in the direction of largely populated areas, for example, Tucson (to the north) and Nogales (to the southwest) areas, and might have a significant impact during observations at low elevations. Since VERITAS simulations are produced for different NSB levels and IRFs are interpolated to match the NSB level of each data run, this effect is corrected to first order in the standard analysis.

\section{Conclusions}

After almost 15 years of operations, VERITAS performance has changed due to a combination of hardware upgrades (relocation of T1, camera and camera electronics upgrades), aging of the different components, and maintenance duties, such as recoating of the most degraded mirror facets. 
With this work, we aim to document the calibration efforts that have been carried out by the VERITAS collaboration during this time. 

As a first step, we described in Section \ref{sec:throughput_measurement} the different approaches that are used to monitor the behavior of the instrument, measuring the gains of each pixel, the relative evolution of the average camera gain over time, and the changes of reflectivity and optical throughput due to telescope aging. The described methodology, now well defined, will continue to be used for the upcoming years of VERITAS operations and we are confident it will serve for other experiments as well.

In Section \ref{sec:throughput_corrections}, we detailed the implementation of the proposed throughput calibration method in our software pipelines. It is based on correction factors for the optical throughput or reflectivity $r_i$ and the average gain of the camera $g_i$, which we combined into a total throughput factor $t_i$. We showed how the application of the throughput factors to the measured signals of the simulated events could be used to produce throughput-calibrated IRFs for VERITAS. These response functions can be used to analyze real showers and derive the corrected particle shower energy and source fluxes. Finally, using the time-dependent response functions obtained, we evaluated the impact of the throughput changes on the performance of VERITAS using different metrics such as the sensitivity and the energy threshold.

In order to validate this calibration method, we performed extensive tests, detailed in Section \ref{sec:validation_realdata}, to ensure that all VERITAS telescopes are properly calibrated against each other and that the geometrical parameters of simulated and real shower events are comparable. In addition, we checked the stability of the flux reconstruction using a multiyear sample of Crab Nebula observations, often assumed to be a reference object in the very-high-energy range.

Finally, in Section \ref{sec:systematic_uncertainties} we discussed the main systematic uncertainties on the energy scale, fluxes, and spectral indices that affect the analysis and reconstruction of data from the VERITAS telescopes. 

As a final note, this work is based on direct measurements of the gains and reflectivity of the dish as an alternative to the study of local muons or the measurement of the cosmic electron spectrum. Both have been proposed and are currently being actively discussed \citep{2016APh....84...23P,2019ApJS..243...11G} as possible calibrators for the upcoming Cherenkov Telescope Array (CTA). CTA will be comprised of arrays several times bigger than that of VERITAS, and have much tighter requirements on systematic uncertainties. It is clear that having independent methods to calibrate the response of the instrument to incident light can only help to reduce their individual limitations and systematic uncertainties.

\begin{acknowledgements}
This research is supported by grants from the U.S. Department of Energy Office of Science, the U.S. National Science Foundation and the Smithsonian Institution, by NSERC in Canada, and by the Helmholtz Association in Germany. This research used resources provided by the Open Science Grid, which is supported by the National Science Foundation and the U.S. Department of Energy's Office of Science, and resources of the National Energy Research Scientific Computing Center (NERSC), a U.S. Department of Energy Office of Science User Facility operated under Contract No. DE-AC02-05CH11231. We acknowledge the excellent work of the technical support staff at the Fred Lawrence Whipple Observatory and at the collaborating institutions in the construction and operation of the instrument.
M.N. acknowledges the Young Investigators Program of the Helmholtz Association for support during the period of the project. Reproduced with permission from Astronomy \& Astrophysics, \copyright ESO.
\end{acknowledgements}

% for the bibliography, at the end
%\begin{thebibliography}
\bibliographystyle{aa} % style aa.bst
\bibliography{bibliography.bib} % your references Yourfile.bib

\begin{thebibliography}{57}
\expandafter\ifx\csname natexlab\endcsname\relax\def\natexlab#1{#1}\fi

\bibitem[{Acciari {et~al.}(2019)Acciari, Ansoldi, Antonelli, Engels, Baack,
  Babi{\'c}, Banerjee, Almeida, Barrio, Gonz{\'a}lez, Bednarek, Bellizzi,
  Bernardini, Berti, Besenrieder, Bhattacharyya, Bigongiari, Biland, Blanch,
  Bonnoli, Bo{\v{s}}njak, Busetto, Carosi, Carosi, Ceribella, Chai,
  Chilingaryan, Cikota, Colak, Colin, Colombo, Contreras, Cortina, Covino,
  D’Amico, D’Elia, Vela, Dazzi, Angelis, Lotto, Delfino, Delgado, Depaoli,
  Pierro, Venere, Espineira, Prester, Donini, Dorner, Doro, Elsaesser,
  Ramazani, Fattorini, Fern{\'a}ndez-Barral, Ferrara, Fidalgo, Foffano,
  Fonseca, Font, Fruck, Fukami, Gallozzi, L{\'o}pez, Garczarczyk, Gasparyan,
  Gaug, Giglietto, Giordano, Godinovi{\'c}, Green, Guberman, Hadasch, Hahn,
  Herrera, Hoang, Hrupec, H{\"u}tten, Inada, Inoue, Ishio, Iwamura, Jouvin,
  Kerszberg, Kubo, Kushida, Lamastra, Lelas, Leone, Lindfors, Lombardi, Longo,
  Lopez, L{\'o}pez-Coto, L{\'o}pez-Oramas, Loporchio, Fraga, Maggio, Majumdar,
  Makariev, Mallamaci, Maneva, Manganaro, Mannheim, Maraschi, Mariotti,
  Mart{\'i}nez, Masuda, Mazin, Mi{\'c}anovi{\'c}, Miceli, Minev, Miranda,
  Mirzoyan, Molina, Moralejo, Morcuende, Moreno, Moretti, Munar-Adrover,
  Neustroev, Nigro, Nilsson, Ninci, Nishijima, Noda, Nogu{\'e}s, N{\"o}the,
  Nozaki, Paiano, Palacio, Palatiello, Paneque, Paoletti, Paredes, Pe{\~n}il,
  Peresano, Persic, Moroni, Prandini, Puljak, Rhode, Rib{\'o}, Rico, Righi,
  Rugliancich, Saha, Sahakyan, Saito, Sakurai, Satalecka, Schmidt, Schweizer,
  Sitarek, {\v{S}}nidari{\'c}, Sobczynska, Somero, Stamerra, Strom, Strzys,
  Suda, Suri{\'c}, Takahashi, Tavecchio, Temnikov, Terzi{\'c}, Teshima,
  Torres-Alb{\`a}, Tosti, Tsujimoto, Vagelli, Scherpenberg, Vanzo, Acosta,
  Vigorito, Vitale, Vovk, Will, Zari{\'c}, \& Nava}]{2020arXiv200607249M}
Acciari, V.~A., Ansoldi, S., Antonelli, L.~A., {et~al.} 2019, Nature, 575, 455

\bibitem[{{Acciari} {et~al.}(2008){Acciari}, {Beilicke}, {Blaylock},
  {Bradbury}, {Buckley}, {Bugaev}, {Butt}, {Byrum}, {Celik}, {Cesarini},
  {Ciupik}, {Chow}, {Cogan}, {Colin}, {Cui}, {Daniel}, {Duke}, {Ergin},
  {Falcone}, {Fegan}, {Finley}, {Fortin}, {Fortson}, {Gall}, {Gibbs},
  {Gillanders}, {Grube}, {Guenette}, {Hanna}, {Hays}, {Holder}, {Horan},
  {Hughes}, {Hui}, {Humensky}, {Kaaret}, {Kieda}, {Kildea}, {Konopelko},
  {Krawczynski}, {Krennrich}, {Lang}, {LeBohec}, {Lee}, {Maier}, {McCann},
  {McCutcheon}, {Millis}, {Moriarty}, {Mukherjee}, {Nagai}, {Ong}, {Pandel},
  {Perkins}, {Pizlo}, {Pohl}, {Quinn}, {Ragan}, {Reynolds}, {Rose},
  {Schroedter}, {Sembroski}, {Smith}, {Steele}, {Swordy}, {Toner}, {Valcarcel},
  {Vassiliev}, {Wagner}, {Wakely}, {Ward}, {Weekes}, {Weinstein}, {White},
  {Williams}, {Wissel}, {Wood}, \& {Zitzer}}]{2008ApJ...679.1427A}
{Acciari}, V.~A., {Beilicke}, M., {Blaylock}, G., {et~al.} 2008, \apj, 679,
  1427

\bibitem[{{Aharonian} {et~al.}(2004){Aharonian}, {Akhperjanian}, {Aye},
  {Bazer-Bachi}, {Beilicke}, {Benbow}, {Berge}, {Berghaus}, {Bernl{\"o}hr},
  {Bolz}, {Boisson}, {Borgmeier}, {Breitling}, {Brown}, {Chadwick}, {Chitnis},
  {Chounet}, {Cornils}, {Costamante}, {Degrange}, {de Jager},
  {Djannati-Ata{\"\i}}, {Drury}, {Ergin}, {Espigat}, {Feinstein}, {Fleury},
  {Fontaine}, {Funk}, {Gallant}, {Giebels}, {Gillessen}, {Goret}, {Guy},
  {Hadjichristidis}, {Hauser}, {Heinzelmann}, {Henri}, {Hermann}, {Hinton},
  {Hofmann}, {Holleran}, {Horns}, {Jung}, {Kh{\'e}lifi}, {Komin}, {Konopelko},
  {Latham}, {Le Gallou}, {Lemoine}, {Lemi{\`e}re}, {Leroy}, {Lohse},
  {Marcowith}, {Masterson}, {McComb}, {de Naurois}, {Nolan}, {Noutsos},
  {Orford}, {Osborne}, {Ouchrif}, {Panter}, {Pelletier}, {Pita}, {Pohl},
  {P{\"u}hlhofer}, {Punch}, {Raubenheimer}, {Raue}, {Raux}, {Rayner},
  {Redondo}, {Reimer}, {Reimer}, {Ripken}, {Rivoal}, {Rob}, {Rolland},
  {Rowell}, {Sahakian}, {Sauge}, {Schlenker}, {Schlickeiser}, {Schuster},
  {Schwanke}, {Siewert}, {Sol}, {Steenkamp}, {Stegmann}, {Tavernet},
  {Th{\'e}oret}, {Tluczykont}, {van der Walt}, {Vasileiadis}, {Vincent},
  {Visser}, {Volk}, \& {Wagner}}]{2004APh....22..109A}
{Aharonian}, F., {Akhperjanian}, A.~G., {Aye}, K.~M., {et~al.} 2004,
  Astroparticle Physics, 22, 109

\bibitem[{{Ajello} {et~al.}(2017){Ajello}, {Atwood}, {Baldini}, {Ballet},
  {Barbiellini}, {Bastieri}, {Bellazzini}, {Bissaldi}, {Blandford}, {Bloom},
  {Bonino}, {Bregeon}, {Britto}, {Bruel}, {Buehler}, {Buson}, {Cameron},
  {Caputo}, {Caragiulo}, {Caraveo}, {Cavazzuti}, {Cecchi}, {Charles},
  {Chekhtman}, {Cheung}, {Chiaro}, {Ciprini}, {Cohen}, {Costantin}, {Costanza},
  {Cuoco}, {Cutini}, {D'Ammando}, {de Palma}, {Desiante}, {Digel}, {Di Lalla},
  {Di Mauro}, {Di Venere}, {Dom{\'\i}nguez}, {Drell}, {Dumora}, {Favuzzi},
  {Fegan}, {Ferrara}, {Fortin}, {Franckowiak}, {Fukazawa}, {Funk}, {Fusco},
  {Gargano}, {Gasparrini}, {Giglietto}, {Giommi}, {Giordano}, {Giroletti},
  {Glanzman}, {Green}, {Grenier}, {Grondin}, {Grove}, {Guillemot}, {Guiriec},
  {Harding}, {Hays}, {Hewitt}, {Horan}, {J{\'o}hannesson}, {Kensei}, {Kuss},
  {La Mura}, {Larsson}, {Latronico}, {Lemoine-Goumard}, {Li}, {Longo},
  {Loparco}, {Lott}, {Lubrano}, {Magill}, {Maldera}, {Manfreda}, {Mazziotta},
  {McEnery}, {Meyer}, {Michelson}, {Mirabal}, {Mitthumsiri}, {Mizuno},
  {Moiseev}, {Monzani}, {Morselli}, {Moskalenko}, {Negro}, {Nuss}, {Ohsugi},
  {Omodei}, {Orienti}, {Orlando}, {Palatiello}, {Paliya}, {Paneque}, {Perkins},
  {Persic}, {Pesce-Rollins}, {Piron}, {Porter}, {Principe}, {Rain{\`o}},
  {Rando}, {Razzano}, {Razzaque}, {Reimer}, {Reimer}, {Reposeur}, {Saz
  Parkinson}, {Sgr{\`o}}, {Simone}, {Siskind}, {Spada}, {Spandre}, {Spinelli},
  {Stawarz}, {Suson}, {Takahashi}, {Tak}, {Thayer}, {Thayer}, {Thompson},
  {Torres}, {Torresi}, {Troja}, {Vianello}, {Wood}, \&
  {Wood}}]{2017ApJS..232...18A}
{Ajello}, M., {Atwood}, W.~B., {Baldini}, L., {et~al.} 2017, \apjs, 232, 18

\bibitem[{{Archambault} {et~al.}(2021){Archambault}, {Chernitsky}, {Griffin},
  \& {Hanna}}]{2021APh...12802556A}
{Archambault}, S., {Chernitsky}, G., {Griffin}, S., \& {Hanna}, D. 2021,
  Astroparticle Physics, 128, 102556

\bibitem[{Archambault {et~al.}(2013)Archambault, Hanna, \&
  Griffin}]{Archambault:2013zja}
Archambault, S., Hanna, D., \& Griffin, S. 2013, in {33rd International Cosmic
  Ray Conference}, 0379

\bibitem[{{Berge} {et~al.}(2007){Berge}, {Funk}, \&
  {Hinton}}]{2007A&A...466.1219B}
{Berge}, D., {Funk}, S., \& {Hinton}, J. 2007, \aap, 466, 1219

\bibitem[{{Bessell}(1990)}]{1990PASP..102.1181B}
{Bessell}, M.~S. 1990, \pasp, 102, 1181

\bibitem[{{Chalme-Calvet} {et~al.}(2014){Chalme-Calvet}, {de Naurois}, \&
  {Tavernet}}]{2014arXiv1403.4550C}
{Chalme-Calvet}, R., {de Naurois}, M., \& {Tavernet}, J.~P. 2014, in {AtmoHEAD
  workshop, 2013}

\bibitem[{{Cogan}(2008)}]{2008ICRC....3.1385C}
{Cogan}, P. 2008, in International Cosmic Ray Conference, Vol.~3, International
  Cosmic Ray Conference, 1385--1388

\bibitem[{{Davies} \& {Cotton}(1957)}]{1957SoEn....1...16D}
{Davies}, J.~M. \& {Cotton}, E.~S. 1957, Solar Energy, 1, 16

\bibitem[{{de Naurois} \& {Mazin}(2015)}]{2015CRPhy..16..610D}
{de Naurois}, M. \& {Mazin}, D. 2015, Comptes Rendus Physique, 16, 610

\bibitem[{Fleury(1991)}]{1991ICRC....2..595F}
Fleury, P. 1991, in International Cosmic Ray Conference, Vol.~2, International
  Cosmic Ray Conference, 595--598

\bibitem[{Garoli {et~al.}(2020)Garoli, Rodriguez De~Marcos, Larruquert, Corso,
  Proietti~Zaccaria, \& Pelizzo}]{app10217538}
Garoli, D., Rodriguez De~Marcos, L.~V., Larruquert, J.~I., {et~al.} 2020,
  Applied Sciences, 10

\bibitem[{{Gaug}(2006)}]{2006PhDT.......205G}
{Gaug}, M. 2006, PhD thesis, Autonomous University of Barcelona

\bibitem[{{Gaug} {et~al.}(2019){Gaug}, {Fegan}, {Mitchell}, {Maccarone},
  {Mineo}, \& {Okumura}}]{2019ApJS..243...11G}
{Gaug}, M., {Fegan}, S., {Mitchell}, A.~M.~W., {et~al.} 2019, \apjs, 243, 11

\bibitem[{{Gazda} {et~al.}(2016){Gazda}, {Nguyen}, {Otte}, \&
  {Richards}}]{2016JInst..1111015G}
{Gazda}, E., {Nguyen}, T., {Otte}, N., \& {Richards}, G. 2016, Journal of
  Instrumentation, 11, P11015

\bibitem[{{Hanna}(2008)}]{2008ICRC....3.1417H}
{Hanna}, D. 2008, in International Cosmic Ray Conference, Vol.~3, International
  Cosmic Ray Conference, 1417--1420

\bibitem[{{Hanna} {et~al.}(2010){Hanna}, {McCann}, {McCutcheon}, \&
  {Nikkinen}}]{2010NIMPA.612..278H}
{Hanna}, D., {McCann}, A., {McCutcheon}, M., \& {Nikkinen}, L. 2010, Nuclear
  Instruments and Methods in Physics Research A, 612, 278

\bibitem[{{Heck} {et~al.}(1998){Heck}, {Knapp}, {Capdevielle}, {Schatz}, \&
  {Thouw}}]{1998cmcc.book.....H}
{Heck}, D., {Knapp}, J., {Capdevielle}, J.~N., {Schatz}, G., \& {Thouw}, T.
  1998, {CORSIKA: a Monte Carlo code to simulate extensive air showers.}
  ({Forschungszentrum Karlsruhe GmbH)})

\bibitem[{{Hillas}(1985)}]{1985ICRC....3..445H}
{Hillas}, A.~M. 1985, in International Cosmic Ray Conference, Vol.~3, 19th
  International Cosmic Ray Conference (ICRC19), Volume 3, 445

\bibitem[{Hillas \& Patterson(1990)}]{Hillas_1990}
Hillas, A.~M. \& Patterson, J.~R. 1990, Journal of Physics G: Nuclear and
  Particle Physics, 16, 1271

\bibitem[{{Hofmann}(2003)}]{2003APh....20....1H}
{Hofmann}, W. 2003, Astroparticle Physics, 20, 1

\bibitem[{{Hofmann} {et~al.}(1999){Hofmann}, {Jung}, {Konopelko},
  {Krawczynski}, {Lampeitl}, \& {P{\"u}hlhofer}}]{1999APh....12..135H}
{Hofmann}, W., {Jung}, I., {Konopelko}, A., {et~al.} 1999, Astroparticle
  Physics, 12, 135

\bibitem[{{Holder}(2005)}]{2005ICRC....5..379H}
{Holder}, J. 2005, in International Cosmic Ray Conference, Vol.~5, 29th
  International Cosmic Ray Conference (ICRC29), Volume 5, 379

\bibitem[{{Holder} {et~al.}(2006){Holder}, {Atkins}, {Badran}, {Blaylock},
  {Bradbury}, {Buckley}, {Byrum}, {Carter-Lewis}, {Celik}, {Chow}, {Cogan},
  {Cui}, {Daniel}, {de la Calle Perez}, {Dowdall}, {Dowkontt}, {Duke},
  {Falcone}, {Fegan}, {Finley}, {Fortin}, {Fortson}, {Gibbs}, {Gillanders},
  {Glidewell}, {Grube}, {Gutierrez}, {Gyuk}, {Hall}, {Hanna}, {Hays}, {Horan},
  {Hughes}, {Humensky}, {Imran}, {Jung}, {Kaaret}, {Kenny}, {Kieda}, {Kildea},
  {Knapp}, {Krawczynski}, {Krennrich}, {Lang}, {LeBohec}, {Linton}, {Little},
  {Maier}, {Manseri}, {Milovanovic}, {Moriarty}, {Mukherjee}, {Ogden}, {Ong},
  {Petry}, {Perkins}, {Pizlo}, {Pohl}, {Quinn}, {Ragan}, {Reynolds}, {Roache},
  {Rose}, {Schroedter}, {Sembroski}, {Sleege}, {Steele}, {Swordy}, {Syson},
  {Toner}, {Valcarcel}, {Vassiliev}, {Wakely}, {Weekes}, {White}, {Williams},
  \& {Wagner}}]{2006APh....25..391H}
{Holder}, J., {Atkins}, R.~W., {Badran}, H.~M., {et~al.} 2006, Astroparticle
  Physics, 25, 391

\bibitem[{{Kieda}(2011)}]{2011ICRC....9...14K}
{Kieda}, D. 2011, in International Cosmic Ray Conference, Vol.~9, International
  Cosmic Ray Conference, 14

\bibitem[{{Kieda}(2013)}]{2013ICRC...33.1124K}
{Kieda}, D.~B. 2013, in International Cosmic Ray Conference, Vol.~33,
  International Cosmic Ray Conference, 1124

\bibitem[{{Krause} {et~al.}(2017){Krause}, {Pueschel}, \&
  {Maier}}]{2017APh....89....1K}
{Krause}, M., {Pueschel}, E., \& {Maier}, G. 2017, Astroparticle Physics, 89, 1

\bibitem[{{Krawczynski} {et~al.}(2006){Krawczynski}, {Carter-Lewis}, {Duke},
  {Holder}, {Maier}, {Le Bohec}, \& {Sembroski}}]{2006APh....25..380K}
{Krawczynski}, H., {Carter-Lewis}, D.~A., {Duke}, C., {et~al.} 2006,
  Astroparticle Physics, 25, 380

\bibitem[{{Krennrich} {et~al.}(2007){Krennrich}, {Blaylock}, {Bradbury},
  {Buckley}, {Byrum}, {Carter-Lewis}, {Celik}, {Cogan}, {Cui}, {Daniel},
  {Dowdall}, {Duke}, {Ergin}, {Falcone}, {Fegan}, {Fortin}, {Fortson}, {Gibbs},
  {Gillanders}, {Gutierrez}, {Grube}, {Hanna}, {Hays}, {Holder}, {Horan},
  {Hughes}, {Humensky}, {Imran}, {Jung}, {Kaaret}, {Kenny}, {Kieda}, {Kildea},
  {Konopelko}, {Krawczynski}, {Lang}, {LeBohec}, {Maier}, {Millis}, {Moriarty},
  {Nagai}, {Ong}, {Perkins}, {Pizlo}, {Quinn}, {Rose}, {Schroedter},
  {Sembroski}, {Smith}, {Steele}, {Swordy}, {Toner}, {Valcarcel}, {Vassiliev},
  {Wagner}, {Wakely}, {Weekes}, {Weinstein}, {White}, {Williams}, \&
  {Wissel}}]{2007JPhCS..60...34K}
{Krennrich}, F., {Blaylock}, G., {Bradbury}, S.~M., {et~al.} 2007, in Journal
  of Physics Conference Series, Vol.~60, Journal of Physics Conference Series,
  34--39

\bibitem[{{Lebohec} \& {Holder}(2003)}]{2003APh....19..221L}
{Lebohec}, S. \& {Holder}, J. 2003, Astroparticle Physics, 19, 221

\bibitem[{{Li} \& {Ma}(1983)}]{1983ApJ...272..317L}
{Li}, T.~P. \& {Ma}, Y.~Q. 1983, \apj, 272, 317

\bibitem[{{L{\'o}pez Paredes} {et~al.}(2018){L{\'o}pez Paredes}, {Ara{\'u}jo},
  {Froborg}, {Marangou}, {Olcina}, {Sumner}, {Taylor}, {Tom{\'a}s}, \&
  {Vacheret}}]{2018APh...102...56L}
{L{\'o}pez Paredes}, B., {Ara{\'u}jo}, H.~M., {Froborg}, F., {et~al.} 2018,
  Astroparticle Physics, 102, 56

\bibitem[{MacLeod(2007)}]{macleod2007measuring}
MacLeod, A. 2007, Master's thesis, McGill University Libraries

\bibitem[{{Maier} \& {Holder}(2017)}]{2017ICRC...35..747M}
{Maier}, G. \& {Holder}, J. 2017, in International Cosmic Ray Conference, Vol.
  301, 35th International Cosmic Ray Conference (ICRC2017), 747

\bibitem[{{Massey} \& {Foltz}(2000)}]{2000PASP..112..566M}
{Massey}, P. \& {Foltz}, C.~B. 2000, \pasp, 112, 566

\bibitem[{{McCann} {et~al.}(2010){McCann}, {Hanna}, {Kildea}, \&
  {McCutcheon}}]{2010APh....32..325M}
{McCann}, A., {Hanna}, D., {Kildea}, J., \& {McCutcheon}, M. 2010,
  Astroparticle Physics, 32, 325

\bibitem[{{Meagher}(2015)}]{2015ICRC...34..792M}
{Meagher}, K. 2015, in International Cosmic Ray Conference, Vol.~34, 34th
  International Cosmic Ray Conference (ICRC2015), 792

\bibitem[{Meyer(2005)}]{2005AIPC..745..774M}
Meyer, M. 2005, in American Institute of Physics Conference Series, Vol. 745,
  High Energy Gamma-Ray Astronomy, ed. F.~A. {Aharonian}, H.~J. {V{\"o}lk}, \&
  D.~{Horns}, 774--778

\bibitem[{{Mirzoyan} {et~al.}(2007){Mirzoyan}, {Garczarczyk}, {Hose}, \&
  {Paneque}}]{2007APh....27..509M}
{Mirzoyan}, R., {Garczarczyk}, M., {Hose}, J., \& {Paneque}, D. 2007,
  Astroparticle Physics, 27, 509

\bibitem[{{Mitchell} {et~al.}(2016){Mitchell}, {Parsons}, {Hofmann}, \&
  {Bernl{\"o}hr}}]{2016APh....75....1M}
{Mitchell}, A.~M.~W., {Parsons}, R.~D., {Hofmann}, W., \& {Bernl{\"o}hr}, K.
  2016, Astroparticle Physics, 75, 1

\bibitem[{{Nagai} {et~al.}(2007){Nagai}, {McKay}, {Sleege}, \&
  {Petry}}]{2007arXiv0709.4517V}
{Nagai}, T., {McKay}, R., {Sleege}, G., \& {Petry}, D. 2007, 30th International
  Cosmic Ray Conference, Merida, Mexico, July 2007, arXiv:0709.4517

\bibitem[{Otte {et~al.}(2011)Otte, Gebremedhin, Kaplan, \&
  Long}]{2011arXiv1110.4702N}
Otte, A.~N., Gebremedhin, L., Kaplan, K., \& Long, D. 2011, in {32nd
  International Cosmic Ray Conference}

\bibitem[{{Parsons} {et~al.}(2016){Parsons}, {Hinton}, \&
  {Schoorlemmer}}]{2016APh....84...23P}
{Parsons}, R.~D., {Hinton}, J.~A., \& {Schoorlemmer}, H. 2016, Astroparticle
  Physics, 84, 23

\bibitem[{{Perkins} {et~al.}(2009){Perkins}, {Maier}, \& {The VERITAS
  Collaboration}}]{2009arXiv0912.3841P}
{Perkins}, J.~S., {Maier}, G., \& {The VERITAS Collaboration}. 2009, 2009 Fermi
  Symposium, eConf Proceedings C091122, arXiv:0912.3841

\bibitem[{Roache {et~al.}(2007)Roache, Irvin, Perkins, Harris, Falcone, Finley,
  \& Weekes}]{roache2007}
Roache, E., Irvin, R., Perkins, J., {et~al.} 2007, in 30th International Cosmic
  Ray Conference (ICRC2017), 1397--1400, 30th International Cosmic Ray
  Conference, ICRC 2007 ; Conference date: 03-07-2007 Through 11-07-2007

\bibitem[{{Roache} {et~al.}(2008){Roache}, {Irvin}, {Perkins}, {Harris},
  {Falcone}, {Finley}, \& {Weeks}}]{2008ICRC....3.1397R}
{Roache}, E., {Irvin}, R., {Perkins}, J.~S., {et~al.} 2008, in International
  Cosmic Ray Conference, Vol.~3, International Cosmic Ray Conference,
  1397--1400

\bibitem[{{Rovero} {et~al.}(1996){Rovero}, {Buckley}, {Fleury}, {Jiang},
  {Pare}, {Sarazin}, {Urban}, \& {Weekes}}]{1996APh.....5...27R}
{Rovero}, A.~C., {Buckley}, J.~H., {Fleury}, P., {et~al.} 1996, Astroparticle
  Physics, 5, 27

\bibitem[{{S{\'a}nchez de Miguel} {et~al.}(2017){S{\'a}nchez de Miguel},
  {Aub{\'e}}, {Zamorano}, {Kocifaj}, {Roby}, \& {Tapia}}]{2017MNRAS.467.2966S}
{S{\'a}nchez de Miguel}, A., {Aub{\'e}}, M., {Zamorano}, J., {et~al.} 2017,
  \mnras, 467, 2966

\bibitem[{Speckmayer {et~al.}(2010)Speckmayer, Höcker, Stelzer, \&
  Voss}]{2007physics...3039H}
Speckmayer, P., Höcker, A., Stelzer, J., \& Voss, H. 2010, Journal of Physics:
  Conference Series, 219, 032057

\bibitem[{{Tyler}(2013)}]{2013ICRC...33.3096T}
{Tyler}, J. 2013, in International Cosmic Ray Conference, Vol.~33,
  International Cosmic Ray Conference, 3096

\bibitem[{{Vacanti} {et~al.}(1994){Vacanti}, {Fleury}, {Jiang}, {Par{\'e}},
  {Rovero}, {Sarazin}, {Urban}, \& {Weekes}}]{1994APh.....2....1V}
{Vacanti}, G., {Fleury}, P., {Jiang}, Y., {et~al.} 1994, Astroparticle Physics,
  2, 1

\bibitem[{{Weekes}(1996)}]{1996SSRv...75....1W}
{Weekes}, T.~C. 1996, \ssr, 75, 1

\bibitem[{Winston(1974)}]{WINSTON197489}
Winston, R. 1974, Solar Energy, 16, 89

\bibitem[{Winston(1976)}]{osti_7129508}
Winston, R. 1976, Light collectors in cylindrical geometry, United States,
  https://www.osti.gov/biblio/7129508

\bibitem[{Zitzer(2013)}]{2013ICRC...33.3076Z}
Zitzer, B. 2013, in International Cosmic Ray Conference, Vol.~33, International
  Cosmic Ray Conference, 3076

\end{thebibliography}
%\end{thebibliography}

\begin{appendix}

\section{Reference mirror reflectivities used in the simulations}\label{sec:appendix_reference_reflectivity}

The reflectivities of the mirror facets used in the simulation of the VERITAS telescopes are based on measurements made in 2011, after T1 was relocated to its current position in the post-upgrade array layout. Together with the average camera gains of section \ref{sec:gains}, they are part of the current detector model of VERITAS, defined in April 2012 and used as a reference throughout this document. The reflectivities were measured from $260\,\mathrm{nm}$ to $700\,\mathrm{nm}$ for several mirrors facets that were located on the top, middle, and bottom of the dish of each telescope. The measured values were then averaged to get a mean reflectivity representative of the entire dish. 

\begin{figure}[!htbp]
\centering\includegraphics[width=0.9\linewidth]{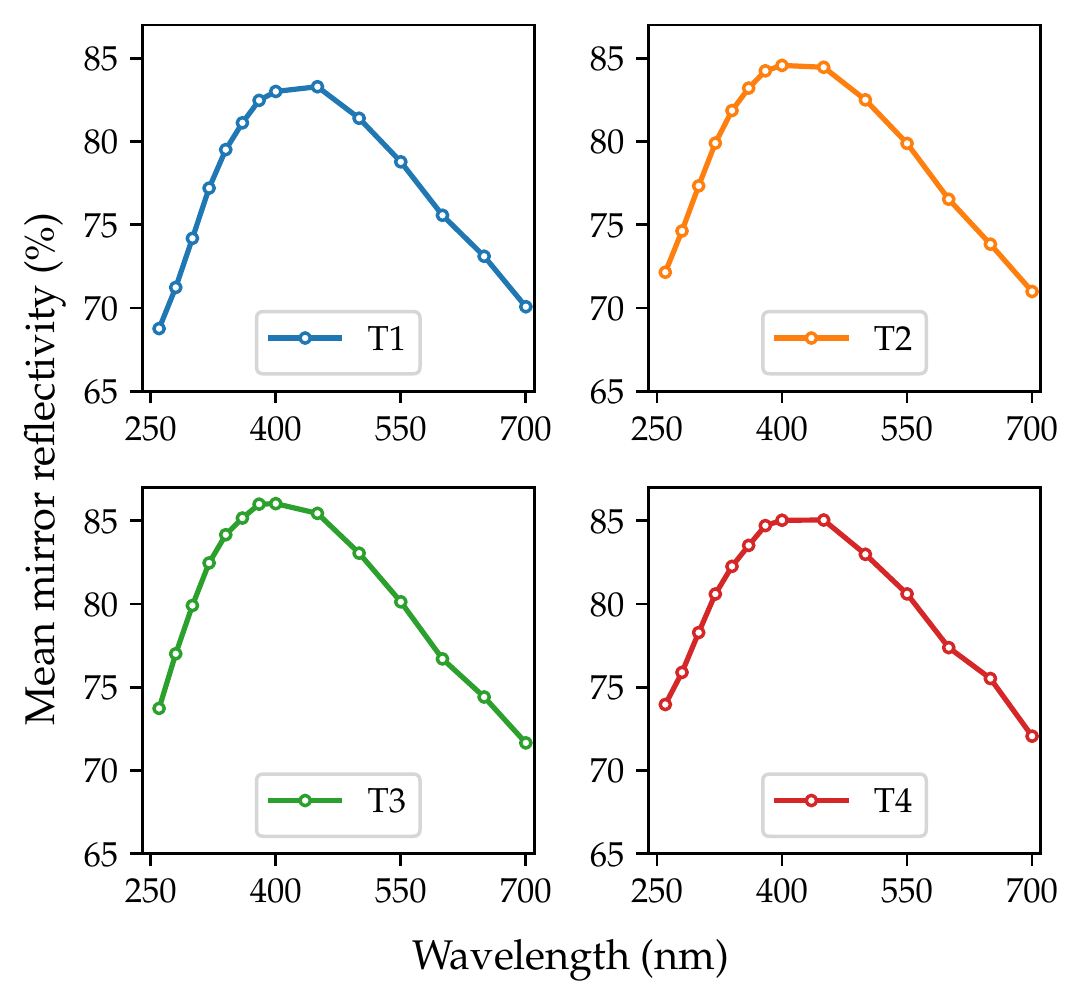}
\caption{Mean telescope reflectivity used to produce the baseline MC simulations in the current instrument configuration of VERITAS, and the reference IRFs discussed throughout this work.}
\label{fig:reference_reflectivity}
\end{figure}

\section{Detailed MC-Data comparison}
\label{sec:appendix_mcdatacomparison}

\begin{figure}[!htbp]
\centering\includegraphics[width=1\linewidth]{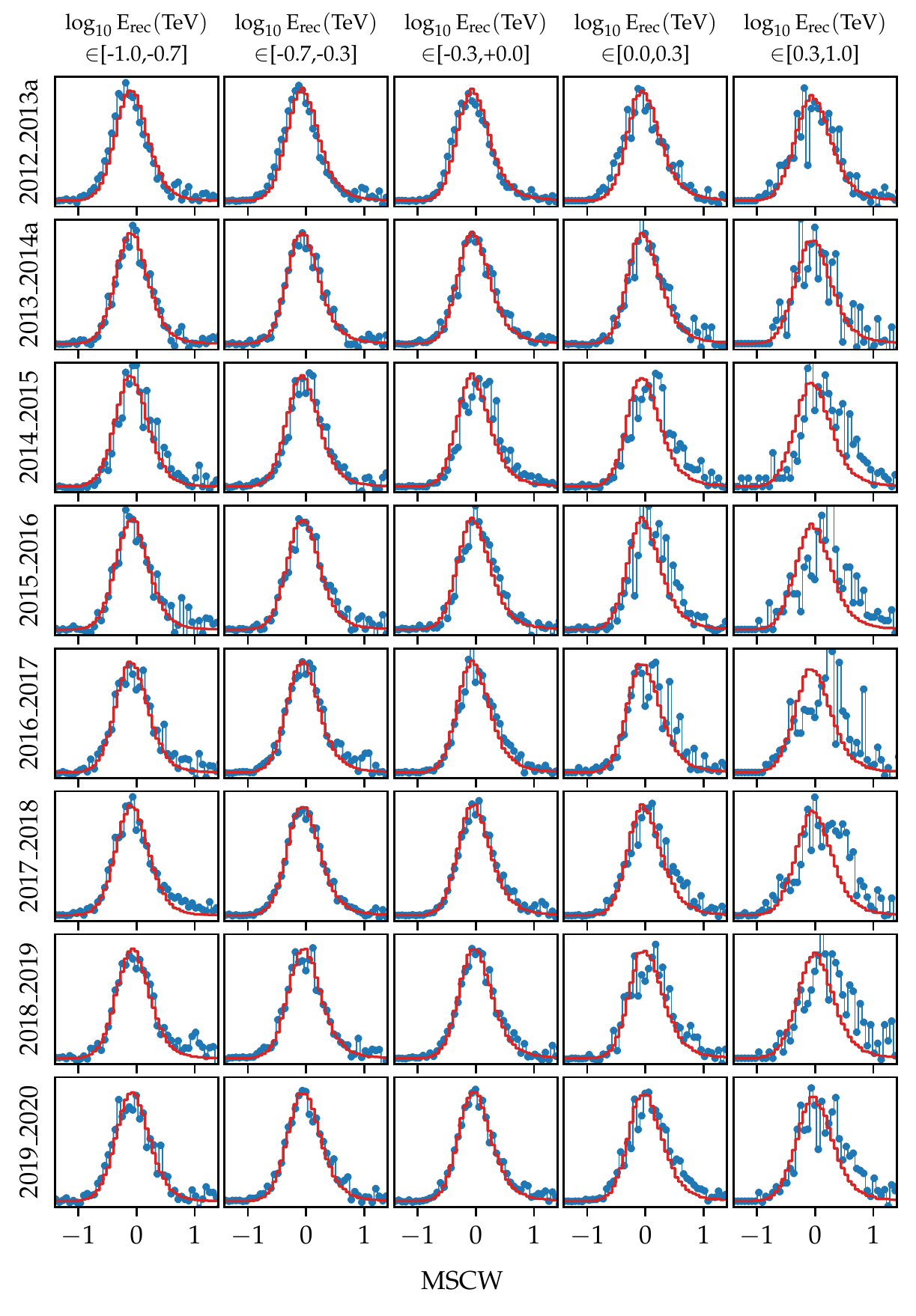}
\caption{Comparison between the MSCW values of simulated $\gamma$-ray showers (red curves) and the ones for real $\gamma$-like events (bliue dots). Each row corresponds to one IRF period and each column to a energy bin (in log-scale, with the energy measured in TeV).}
\label{fig:mscw_data_mc_period}
\end{figure}

\begin{figure}[!htbp]
\centering\includegraphics[width=1\linewidth]{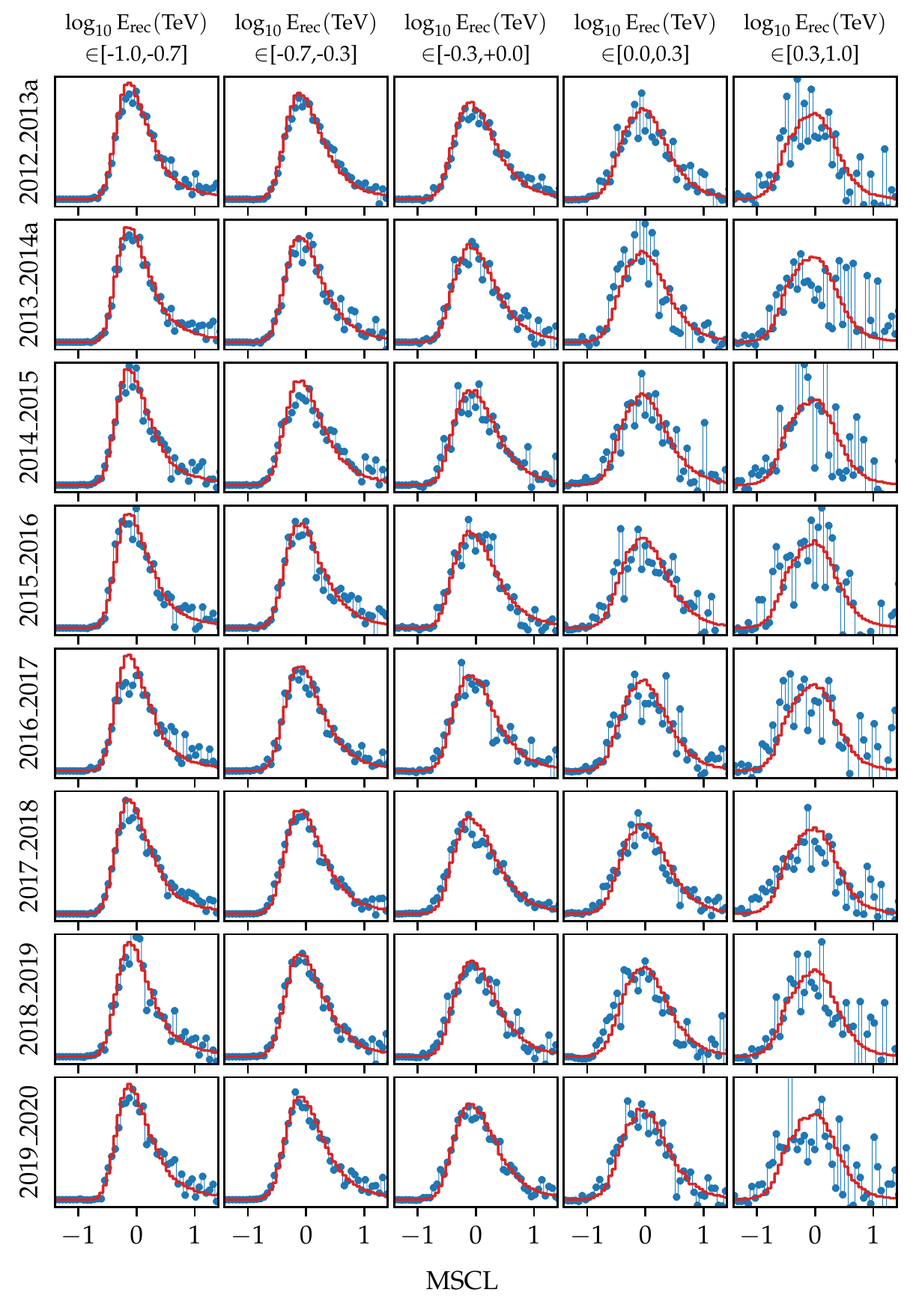}
\caption{Comparison between the MSCL values of simulated $\gamma$-ray showers (red curves) and the ones for real $\gamma$-like events (blue dots). Each row corresponds to one IRF period and each column to a energy bin (in log-scale, with the energy measured in TeV).}
\label{fig:mscl_data_mc_period}
\end{figure}

During the validation of the proposed method to calibrate the throughput, we performed extensive tests to check the agreement between data and the modified MC simulations. One of the most sensitive tests in VERITAS to detect possible discrepancies between the measured Cherenkov signals and the corresponding simulated showers is to check the agreement in the MSCW and MSCL parameters. Both are shown, for each IRF period, in Figures \ref{fig:mscw_data_mc_period} and \ref{fig:mscl_data_mc_period}.

\section{Stability of spectral parameters}

As discussed in section \ref{sec:VTS_performance}, the Crab Nebula is often used as a source to benchmark the performance of $\gamma$-ray instruments. It is not only one of the brightest sources in the sky, but also visible from both hemispheres.
The Crab Nebula has a hard spectrum which reaches energies of a few tens of TeV with photon fluxes that are detectable by VERITAS. Finally, it is thought to be stable in year-timescales both in spectral shape and emitted luminosity, at least in the very-high-energy band. Using eight one-year observation campaigns of the Crab Nebula for the entire post-upgrade period of VERITAS, we could evaluate the performance of the proposed throughput calibration method.

\begin{figure}[!htbp]
\centering\includegraphics[width=1.0\linewidth]{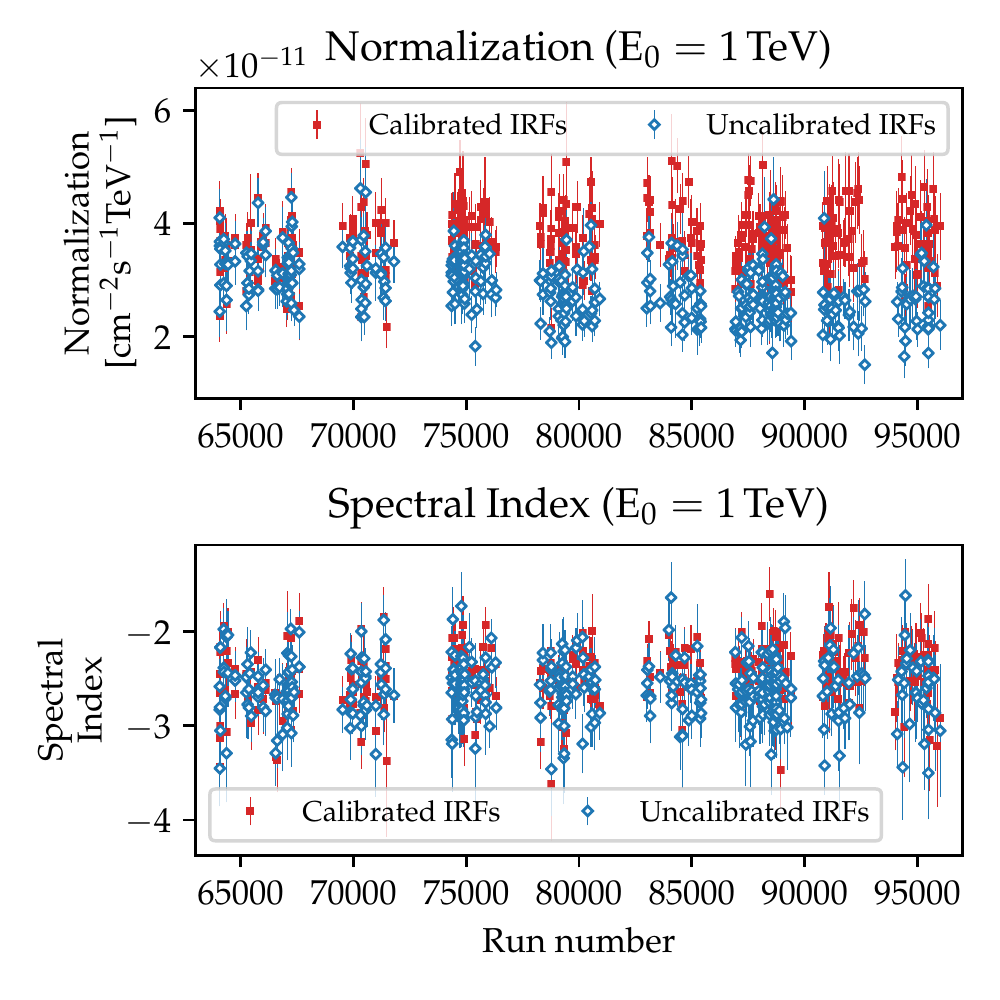}
\caption{Run-wise measurement of the normalization flux at $1\,\mathrm{TeV}$ and the spectral index of the Crab Nebula between 0.6 and 4 TeV, including all IRF periods starting from the upgrade of VERITAS. Only runs with a detection significance of at least $4\,\sigma$ are included.}
\label{fig:energy_spectra_crab_time}
\end{figure}

\begin{figure}[!htbp]
\centering\includegraphics[width=1.0\linewidth]{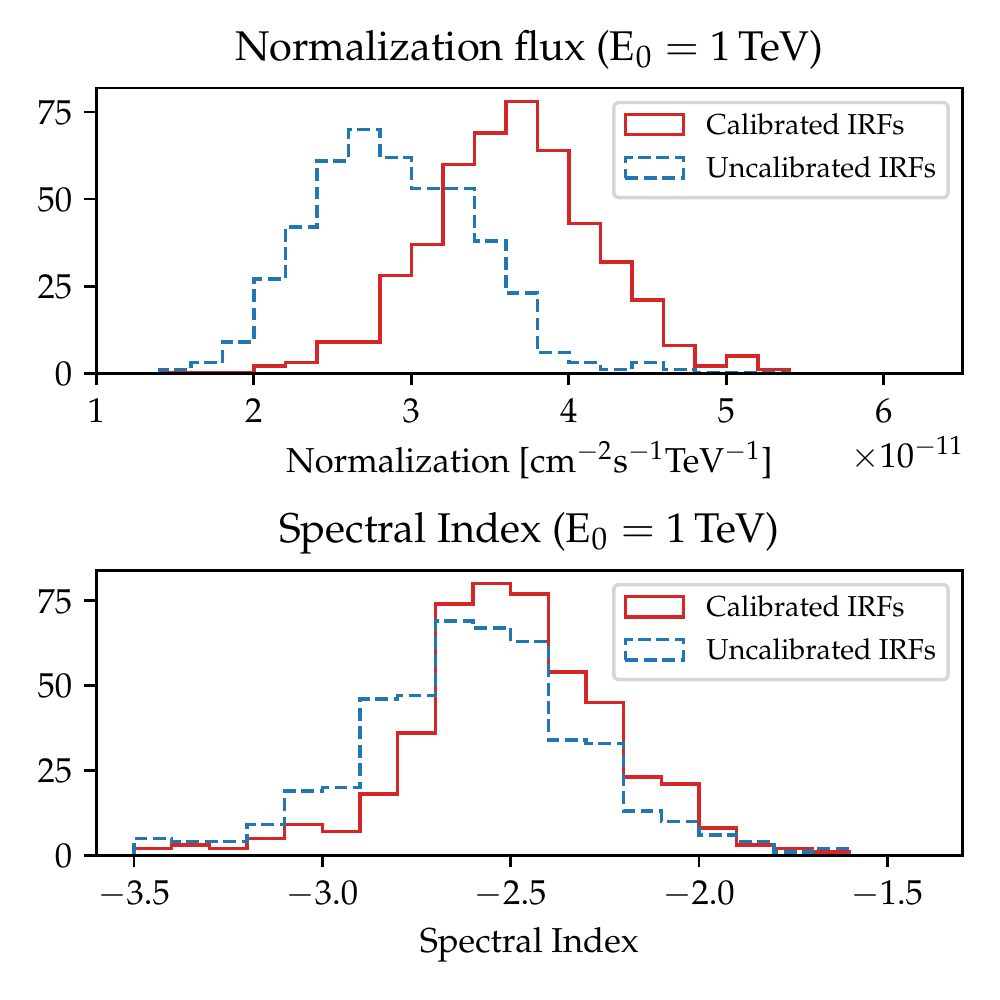}
\caption{Histogram of run-wise measurements of normalization and spectral index of Crab Nebula between 0.6 and 4 TeV including all IRF periods starting from the upgrade of VERITAS. Only runs with a detection significance of at least $4\,\sigma$ are included.}
\label{fig:energy_spectra_crab_hist}
\end{figure}

Figure \ref{fig:energy_spectra_crab_time} shows as a time series the run-wise values of the normalization flux and the spectral index of the source, evaluated at $1\,\mathrm{TeV}$. It was estimated by analysing each Crab Nebula run that survived quality cuts and had more than $4\,\sigma$ excess. Then, the resulting run-wise spectra were fitted locally with a power-law over the small energy range of $0.4-6\,\mathrm{TeV}$. A histogram representation of the values of Figure \ref{fig:energy_spectra_crab_time} can be seen in Figure \ref{fig:energy_spectra_crab_hist}.  

\end{appendix}

%-------------------------------------------------------------------

\end{document}